\documentclass[a4paper,12pt]{article}
\usepackage{jheppub} 
\usepackage{braket}
\usepackage{makecell}
\usepackage{diagbox}
\usepackage{subfigure}

\newcommand\td{\text{d}}
\newcommand{\p}{\partial}

\renewcommand{\thefootnote}{\fnsymbol{footnote}} 
\def\>{\rangle} \def\<{\langle}
\title{\boldmath \Large Character of the highest weight module of BMS algebra realized on codimensional-two boundary} 
\author[a,b,c]{Bin Chen$^*$}
\author[a,e]{,\,Song He$^*$}
\author[d]{,\,Pujian Mao$^*$}
\author[b]{,\,Xin-Cheng Mao$^*$\footnote{$^*$: Corresponding authors}}

\affiliation[a]{Institute of Fundamental Physics and Quantum Technology, \\\&  School of Physical Science and Technology, \\ Ningbo University, Ningbo, Zhejiang 315211, China}
\affiliation[b]{School of Physics, Peking University, \\No.5 Yiheyuan Rd, Beijing 100871, P.~R.~China}
\affiliation[c]{Center for High Energy Physics, Peking University, \\No.5 Yiheyuan Rd, Beijing 100871, P.~R.~China}
\affiliation[d]{ Center for Joint Quantum Studies, Department of Physics, School of Science, \\Tianjin University, 135 Yaguan Road, Tianjin 300350, China}
\affiliation[e]{Max Planck Institute for Gravitational Physics (Albert Einstein Institute), \\ Am M$\Ddot{\text u}$hlenberg 1, 14476 Golm, Germany}

\emailAdd{chenbin1@nbu.edu.cn,\, hesong@nbu.edu.cn,\, pjmao@tju.edu.cn,\, maoxc1120@stu.pku.edu.cn}

\abstract{In this paper, we investigate the highest weight representation (HWR) of the three and four-dimensional Bondi-Metzner-Sachs (BMS) algebra realized on the codimension-two boundary of asymptotic flat spacetime (AFS). In such a realization, the action of supertranslation shifts the conformal weight of the highest-weight states. As a result, there is no extra quantum number relating to the supertranslation. We construct the highest-weight BMS modules and compute their characters. We show that the BMS$_3$ highest-weight vacuum character with special value of central charges coincides with the 1-loop partition function of three-dimensional asymptotic flat gravity, up to an overall phase factor ``$i$''. We expect the vacuum character of BMS$_4$ may shed light on the flat holography in four dimensions.}

\begin{document}
\maketitle
\flushbottom

\renewcommand{\thefootnote}{\arabic{footnote}} 

\section{Introduction} \label{sec:intro}

Symmetry plays a fundamental role in the formulation of physical theories. The most renowned example of an application of symmetry in physics is Noether’s theorem \cite{Noether:1918zz}, which states that a global continuous symmetry leads to a conservation current. In modern quantum field theory and particle physics,  Lorentz symmetry and gauge symmetry play indispensable roles. For example, the physical fields or their excitations are classified by their representations of the Lorentz group. Symmetry also plays a fundamental role in gravity.  Over the past two decades, representation theories of symmetry algebras have attracted significant attention in the study of three-dimensional gravity from the holographic perspective, applicable to both asymptotically Anti-de Sitter (AdS) and flat spacetimes. The fuel booster source for the recent activity is from the mathematically exact dual relation between the one-loop partition function of free linearized metric perturbation on AdS space or flat space and the character of the vacuum representation of a symmetry algebra \cite{Witten:2007kt,Maloney:2007ud,Giombi:2008vd,David:2009xg,Barnich:2015mui,Oblak:2015sea}. The surprising fact is that the symmetry algebra is not the isometry of the background spacetime but the asymptotic symmetry algebra, i.e., two copies of the Virasoro algebra \cite{Brown:1986nw} for the AdS$_3$ case and the BMS$_3$ algebra for the flat case \cite{Ashtekar:1996cd, Barnich:2006av}.

The asymptotic symmetries are those gauge transformations that leave the field configurations under consideration asymptotically invariant. They were first introduced in the investigation of gravitational radiations in four-dimensional Einstein theory, and they are now referred to as Bondi-Metzner-Sachs (BMS) symmetries \cite{Bondi:1962px,Sachs:1962wk,Sachs:1962zza}. In four dimensions, the BMS symmetry algebra consists of the semidirect sum of an infinite-dimensional supertranslation that extends the Poincar\'{e} translations and the Lorentz transformations.
Recently, Barnich and Troessaert have shown that the Lorentz part in the original BMS$_4$ symmetry \cite{Bondi:1962px,Sachs:1962wk,Sachs:1962zza} could be enlarged to local conformal symmetries of the celestial sphere, namely, two copies of the infinite-dimensional Witt algebra, by removing the constraints from global considerations \cite{Barnich:2009se,Barnich:2010ojg,Barnich:2010eb}. We would refer to the Barnich-Troessaert version of BMS$_4$ symmetry as a local version of BMS$_4$ symmetry, or extended BMS$_4$ symmetry. The appearance of Witt algebra suggests that the techniques in two-dimensional conformal field theory (CFT) will play a fundamental role, particularly from a holographic perspective. Indeed, the discovery in \cite{Barnich:2009se,Barnich:2010ojg,Barnich:2010eb} has inspired recent intensive investigations into the applications of asymptotic symmetries; see, e.g., the comprehensive review in \cite{Strominger:2017zoo}. All these recent efforts about asymptotic symmetries converge on building the flat holography in four-dimensional spacetime from the celestial perspective \cite{Pasterski:2016qvg, Pasterski:2017kqt, Pasterski:2017ylz, Raclariu:2021zjz, Pasterski:2021rjz, Pasterski:2021raf} and the Carrollian perspective \cite{Donnay:2022aba, Donnay:2022wvx, Bagchi:2022emh, Chen:2023naw, Chen:2023pqf, Bagchi:2023cen}.

In this paper, we will focus on the highest-weight representations of the local BMS$_4$ symmetry \cite{Barnich:2009se,Barnich:2010ojg,Barnich:2010eb},\footnote{The unitary irreducible representations of the global version of the BMS$_4$ symmetry have been studied in, e.g., \cite{RLINA_1967_8_43_1-2_a5, cantoni1966class, cantoni1967reduction, Crampin:1974aw, Girardello:1974sq, mccarthy1, mccarthy2, mccarthy3, mccarthy4, Mccarthy:1972ry, mccarthy1978lifting, mccarthy1975bondi, mccarthy78hyp,Bekaert:2024uuy, Bekaert:2025kjb, Prabhu:2022zcr, Prabhu:2024lmg, Prabhu:2024zwl}, also \cite{Barnich:2021dta,Barnich:2022bni,AliAhmad:2025hdl} for recent progress with extensions to the local version of the BMS$_4$ symmetry.} which is inspired by the remarkable applications of symmetry algebra \cite{Witten:2007kt,Maloney:2007ud,Giombi:2008vd,David:2009xg,Barnich:2015mui,Oblak:2015sea} to three-dimensional holography. 
We study the highest weight representation (HWR) of the BMS$_4$ algebra, which is realized on the codimension-two boundary, the celestial sphere $S^2$. The motivation for a codimension-two realization stems from the ongoing four-dimensional flat holography, which connects the scattering amplitudes of the four-dimensional bulk theory to the correlators of the CFT on the celestial sphere.

As a heuristic warm-up, in section \ref{sec:BMS3}, we start from the HWR of the BMS$_3$ algebra \cite{Ashtekar:1996cd, Barnich:2006av} through the BMS$_3$ invariant field theory on a circle (the codimension-two boundary $S^1$). 
Each HWR is constructed from a unique highest-weight state, and its descendants are generated from the elements of the BMS algebra.We then compute the character of the highest-weight BMS$_3$ modules. It turns out that the supertranslation and superrotation sectors are factorized, each written in terms of the Euler function. 

It is well-known that three-dimensional Einstein gravity has no local propagating degree of freedom. In three-dimensional cases, the bulk gravity theory is entirely determined by the boundary dynamics. In particular, for a negative cosmological constant, the one-loop gravitational partition function around the pure AdS saddle point exactly matches the character of the highest-weight vacuum module of the two-dimensional conformal group (Virasoro$\times$Virasoro) \cite{Witten:2007kt,Maloney:2007ud,Giombi:2008vd,David:2009xg}. For a vanishing cosmological constant, the one-loop gravitational partition function around the Minkowski saddle point is precisely equivalent to the character of the induced vacuum module of BMS$_3$ \cite{Barnich:2015mui,Oblak:2015sea}, which is an unitary representation\footnote{For other investigations on the unitary irreducible representations of BMS$_3$, see, e.g., \cite{Barnich:2014kra,Barnich:2015uva, Oblak:2016eij, Campoleoni:2016vsh, Melas:2017ggm, Melas:2017jzb, Melas:2021oje}. In contrast, in our novel constructions of HWRs realized on the codimensional-two boundary, the Hilbert space inevitably contains zero-norm states, since the supertranslation descendants are zero-norm states but do not satisfy the same highest-weight conditions as the primary states or the highest-weight vacuum.
The character of the highest-weight vacuum module from our construction matches the one-loop gravitational partition function up to an overall phase factor ``$i$'' with a special choice of the central charge of the BMS$_3$ algebra, which is different from the choice in the induced representation \cite{Barnich:2015mui,Oblak:2015sea}.}

Due to the structure of the BMS$_3$ group, the supertranslation can only act as a conformal weight shifting operator \cite{Stieberger:2018onx, Fotopoulos:2019vac} on $S^1$, and thus the primary states in this case can only be characterized by a single quantum number—the conformal weight. By contrast, the supertranslation can act as the coordinate transformation in higher-dimensional manifolds, such as two-dimensional Carrollian manifold or Galilean manifold \cite{Hao:2021urq, Hao:2022xhq, Saha:2022gjw, Chen:2020vvn, Chen:2022jhx, Chen:2024voz, Bagchi:2009pe, Bagchi:2009ca, Bagchi:2016geg, Bagchi:2019unf, Chen:2021xkw}, it leads to more complicated structures: the primary states are characterized by both the scaling dimension and boost charge in Jordan form, allowing for the emergence of multiplet structures. The HWR realized on $S^1$ corresponds to the singlet version with fixed boost charge.

We then perform a parallel investigation for the BMS$_4$ algebra in section \ref{sec:BMS4}. 
Similar to the BMS$_3$ case, the supertranslation action only shifts the conformal weight of the highest weight states on $S^2$. However, it shifts the conformal weight by half-integers, and there is no generator of supertranslation commuting with $L_0, \bar{L}_0$. Consequently, there is no quantum number for a primary state with respect to the supertranslation in this case. In contrast, supertranslations are more intricate on a 3D Carrollian manifold than on $S^2$, the highest weight states on the Carrollian manifold exhibit generically multiplet structures \cite{Chen:2021xkw}, while the HWRs realized on $S^2$ present only a much simpler singlet structure. This leads to significant consequences in the computation of the character of the highest-weight BMS$_4$ modules.

In the computation of the character of the highest-weight BMS$_4$ modules realized on $S^2$, the supertranslation sector is also factored out from the superrotation part, but its contribution cannot be expressed as an Euler function. Fortunately, it can be obtained in a remarkably compact form by applying other generating functions of partitions. To our knowledge, this is the very first computation of a character from the representation theory of the local version of the BMS$_4$ algebra. The potential applications can be very fruitful. 
As the gravity theory in four dimensions has a local propagating degree of freedom, it is reasonable to expect a gravitational dual of both the highest weight states and the highest weight vacuum of the BMS$_4$ invariant field theory on the codimension-two boundary. Our results for the BMS$_4$ characters can serve as a stepping stone toward a better understanding of flat holography from the perspective of the partition function.


\section{Highest weight representation of BMS$_3$ realized on $S^1$} \label{sec:BMS3}

The structure of BMS$_3$ group is a semidirect product of the superrotations, which is the diffeomorphism on the one-dimensional circle $S^1$, and supertranslations, arbitrary smooth functions on $S^1$:
\begin{equation} \label{eq:BMS3stru}
    \mathrm{BMS}_3 = \mathrm{Diff} (S^1) \ltimes C^{\infty} (S^1).
\end{equation}
In this section, we study the BMS$_3$-invariant field theory on $S^1$, starting from the centrally extended BMS$_3$ algebra \cite{Ashtekar:1996cd, Barnich:2006av, Barnich:2010ojg,Barnich:2012xq, Barnich:2012aw}:
\begin{equation} \label{eq:cBMS3A}
    \begin{split}
        [L_n, L_m] & = (n-m) L_{n+m} + \frac{c_L}{12} (n^3 - n) \delta_{n+m,0}, \\
        [L_n, M_m] & = (n-m) M_{n+m} + \frac{c_M}{12} (n^3 - n) \delta_{n+m,0}, \\
        [M_n, M_m] & = 0,
    \end{split}
\end{equation}
where $L_n$ and $M_n$ generate superrotations and supertranslations, respectively. Here, the Virasoro generators $L_n$ generate local conformal transformations on the celestial sphere $S^1$. Consequently, the resulting theory is a one-dimensional conformal field theory with additional supertranslation symmetries\footnote{In fact, the term supertranslation was originally introduced in the context of AFS$_3$ at null infinity. For simplicity, in this paper, whenever we refer to ``supertranslations'' or ``superrotations'', we mean their corresponding elements within the BMS group or BMS algebra.}. 

Note that the canonical realizations of BMS transformations depend on the manifolds on which they act. Therefore, field theories sharing the same BMS symmetry but defined on different manifolds may be formally distinct. Another extensively studied BMS$_3$-invariant field theory is defined on a 2D Carrollian manifold, commonly referred to as 2D Carrollian CFT (CarCFT$_2$) \cite{Hao:2021urq, Hao:2022xhq, Saha:2022gjw, Chen:2020vvn, Chen:2022jhx, Chen:2024voz, Bagchi:2009pe, Bagchi:2009ca, Bagchi:2016geg, Bagchi:2019unf, Chen:2021xkw}. In contrast to the CarCFT$_2$ where the supertranslation can generate boost charge, the supertranslation can only act as an operator that shifts the conformal weight of the state \cite{Stieberger:2018onx, Fotopoulos:2019vac} on $S^1$, because the superrotations encompass all coordinate transformations on $S^1$, based on the group structure \eqref{eq:BMS3stru}. 

The goal of this section is to construct the highest-weight representations of BMS$_3$ within the framework of the BMS$_3$ invariant field theory on $S^1$,
and to compute the corresponding highest-weight characters. In particular, we will elucidate the relationship between the character of the highest-weight representation generated by the vacuum state and the bulk gravity partition function.

\subsection{BMS$_3$ transformations and OPEs in  field theory on $S^1$} \label{sec:OPE1d}

A circle $S^1$ is topologically equivalent to the compactified real line $\mathbb{R} \cup \{\infty\}$, which can be parametrized by the coordinate $z = \cot \frac{\phi}{2}$, where the angular coordinate $\phi$ is periodic with period $2\pi$. Throughout this section, we work on the variable $z$. We will encounter real one-dimensional integrals containing poles on the real axis. These are evaluated via principal value prescriptions and the residue theorem, which requires promoting $z$ to a complex variable.

To incorporate all BMS$_3$ modes, we define the quasi-primary currents:
\begin{equation} \label{eq:TMmode1d}
    T(z) = \sum_{n \in \mathbb{Z}} z^{-n-2} L_n, \qquad M(z) = \sum_{n \in \mathbb{Z}} z^{-n-2} M_n,
\end{equation}
where $T(z)$ is the stress tensor generating conformal (superrotation) transformations, and $M(z)$ is the current generating supertranslations. The BMS$_3$ generators can be recovered as:
\begin{equation}
    L_n = \oint_0 \frac{\td z}{2\pi i} z^{n+1} T(z), \qquad M_n = \oint_0 \frac{\td z}{2\pi i} z^{n+1} M(z).
\end{equation}
These mode operators act as BMS$_3$ charges. While individual modes capture part of the symmetry, the full BMS$_3$ transformation should account for total charges that contain the contributions of all modes. The total charge can be constructed from the currents $T, M$ through the residue theorem
\begin{equation} \label{eq:ModeQS1}
\begin{aligned}
    Q_\varepsilon &= \oint \frac{\td z}{2\pi i} T(z) \varepsilon(z) = \sum_{n \in \mathbb{Z}} \varepsilon_n L_n, \\
    Q_\beta &= \oint \frac{\td z}{2\pi i} M(z) \beta(z) = \sum_{n \in \mathbb{Z}} \beta_n M_n,
\end{aligned}
\end{equation}
with mode expansions:
\begin{equation}
    \varepsilon(z) = \sum_{n \in \mathbb{Z}} z^{n+1} \varepsilon_n, \qquad \beta(z) = \sum_{n \in \mathbb{Z}} z^{n+1} \beta_n.
\end{equation}
The full BMS$_3$ transformation of the fields is then generated by their commutators with $Q_\varepsilon, Q_\beta$. Before analyzing the BMS$_3$ transformation, it is useful first to examine the operator product expansion (OPE) between currents and operators, as the OPEs are related to commutators by counting the residues
\begin{equation} \label{eq:[A,b]}
    [A, b(w)] = \oint_w \frac{\td z}{2\pi i} a(z) b(w), \qquad [A, B] = \oint_0 \frac{\td w}{2\pi i} \oint_w \frac{\td z}{2\pi i} a(z) b(w),
\end{equation}
for operators $A = \oint_0 \frac{\td z}{2\pi i} a(z)$ and $B = \oint_0 \frac{\td w}{2\pi i} b(w)$. 
For example, the OPEs between the currents read as follows from the charge algebra \eqref{eq:cBMS3A}
\begin{equation} \label{eq:TMOPE}
\begin{aligned}
    T(z) T(w) &\sim \frac{c_L/2}{(z-w)^4} + \frac{2 T(w)}{(z-w)^2} + \frac{\partial_w T(w)}{z-w}, \\
    T(z) M(w) &\sim \frac{c_M/2}{(z-w)^4} + \frac{2 M(w)}{(z-w)^2} + \frac{\partial_w M(w)}{z-w}, \\
    M(z) M(w) &\sim 0.
\end{aligned}
\end{equation}
Under a finite conformal transformation $z \mapsto z'(z)$, the quasi-primary currents transform as:
\begin{equation} \label{eq:finite1DTM}
\begin{aligned}
    T'(z') &= \left(\frac{\td z'}{\td z}\right)^{-2} \left[ T(z) - \frac{c_L}{12} \{z'; z\} \right], \\
    M'(z') &= \left(\frac{\td z'}{\td z}\right)^{-2} \left[ M(z) - \frac{c_M}{12} \{z'; z\} \right],
\end{aligned}
\end{equation}
where the Schwarzian derivative is defined by:
\begin{equation}
    \{F(x); x\} \equiv \frac{F'''(x)}{F'(x)} - \frac{3}{2} \left( \frac{F''(x)}{F'(x)} \right)^2.
\end{equation}

We define a ``primary field'' $\phi_h (z)$ that is characterized by the following behavior under BMS$_3$ transformation
\begin{equation} \label{eq:HWS1}
    [L_n, \phi_h (z)] = l_n \phi_h (z), \qquad [M_n, \phi_h (z)] = m_n \phi_h (z),
\end{equation}
where
\begin{equation} \label{eq:Virhwrep}
    l_n = -z^{n+1} \partial_z - (n+1) z^n h, \qquad m_n = z^{n+1} e^{\partial_h}.
\end{equation} 
This reflects the canonical BMS$_3$ transformation through the primary field $\phi_h$, as $l_n, m_m$ satisfies the BMS$_3$ algebra ($c_L = c_M = 0$ in \eqref{eq:cBMS3A})\footnote{To verify the BMS$_3$ algebra, one needs to apply the relation $[e^{\p_h},h]=e^{\p_h}$.}

The OPEs between the currents and a conformal primary field $\phi_h(z)$ follow from these commutators:
\begin{equation} \label{eq:TMphiOPE}
\begin{aligned}
    T(z) \phi_h(w) &\sim \frac{h \phi_h(w)}{(z - w)^2} + \frac{\partial_w \phi_h(w)}{z - w}, \\
    M(z) \phi_h(w) &\sim \frac{\phi^M_{h+1}(w)}{z - w}, 
\end{aligned}
\end{equation}
where $\phi^M_{h+1}(z) = e^{\partial_h} \phi_h(z)$ is a weight-$(h+1)$ operator similar to, but slightly different from the primary operator $\phi_{h+1} (z)$. The properties of $\phi^M_{h+1}(z)$ will be discussed in detail later. The OPEs imply Ward identities for the correlator functions:
\begin{equation} \label{eq:TMWardS1}
\begin{aligned}
    \langle T(z) X_n \rangle &= \sum_{k=1}^n \left[ \frac{h_k}{(z - z_k)^2} + \frac{\partial_{z_k}}{z - z_k} \right] \langle X_n \rangle, \\
    \langle M(z) X_n \rangle &= \sum_{k=1}^n \frac{e^{\partial_{h_k}}}{z - z_k} \langle X_n \rangle,
\end{aligned}
\end{equation}
where $X_n = \prod_{i=1}^n \phi_{h_i}(z_i)$. 
The currents $T(z)$ and $M(z)$ are quasi-primary due to the fourth-order poles in their OPE \eqref{eq:TMOPE}. 

The infinitesimal BMS$_3$ transformation on primary operators can be generated from the charges \eqref{eq:ModeQS1} as
\begin{equation} \label{eq:BMS3transS1}
\begin{split}
    [Q_\varepsilon, \phi_h(z)] &= \left[ h \partial_z \varepsilon(z) + \varepsilon(z) \partial_z \right] \phi_h(z), \\
    [Q_\beta, \phi_h(z)] &= \beta(z) \phi^M_{h+1}(z),
\end{split}
\end{equation}
which provides an equivalent definition of the BMS$_3$ representation given in \eqref{eq:Virhwrep}. The finite version of conformal transformation $z \mapsto z'(z)$ of primary fields is:
\begin{equation} \label{eq:1dconphi}
    \phi'_h(z') = \left( \frac{\td z'}{\td z} \right)^{-h} \phi_h(z).
\end{equation}
Then, one can clarify that $\phi^M_{h-n}(z) = [M_n, \phi_h (z)]$ does not transform as a weight-$(h-n)$ conformal primary field unless $n=0$, which gives a first glance at the role of supertranslations. The field $\phi^M_{h+1}$ introduced before is a special case of this catalog. This is evident from the analysis of the mode expansions. The primary field has the expansion:
\begin{equation} \label{eq:modephi1d}
    \phi_h(z) = z^{-h} \sum_{n \in \mathbb{Z}} z^{-n} \phi^h_n.
\end{equation}
The mode expansion of $\phi^M_{h-n} (z)$ can then be derived by acting $z^{n+1} e^{\p_h}$ on the primary operator $\phi_h (z)$ as
\begin{equation}
    \phi^M_{h-n}(z) = z^{-h+n} \sum_{m \in \mathbb{Z}} z^{-m} \phi^h_m = z^n \phi_h (z) ,
\end{equation}
where we used $e^{\p_h} z^{-h} = z^{-h - 1}$. Therefore, supertranslations shift the weight in general when $ n\neq 0$, but the resulting field is not a primary field unless $n=0$. This can be observed from its behavior under a finite conformal transformation 
\begin{equation}
    \phi^{M \prime}_{h-n} (z') = z^{\prime n} \phi'_h (z') = \left( \frac{z'}{z} \right)^n \left( \frac{\td z'}{\td z} \right)^{-h} \phi'_h (z').
\end{equation}
A more detailed description will be discussed in the following subsection.

\subsection{Asymptotic states}

The Hilbert spaces of in- and out-states are defined at $z = 0$ and $z = \infty$, respectively, and are related through Hermitian conjugation. Before determining the precise conjugation structure, we first define the in- and out-states independently. Leveraging the state-operator correspondence in conformal field theory, the primary in-states are constructed by inserting primary operators at the origin on the vacuum state $\ket{\text{vac}}$:
\begin{equation} \label{eq:|h>1d}
    \ket{h} = \lim_{z \rightarrow 0} \phi_h(z) \ket{\text{vac}}.
\end{equation}
To ensure convergence at $z = 0$, modes that create singularities must annihilate the vacuum. From the mode expansion \eqref{eq:modephi1d}, this implies the condition:
\begin{equation} \label{eq:anniphi1d}
    \phi_n^h \ket{\text{vac}} = 0 \qquad \text{for} \quad n > -h.
\end{equation}
The dual bra state is defined via Hermitian conjugation:
\begin{equation}
    \bra{h} = \ket{h}^{\dagger} = \lim_{z \rightarrow 0} \bra{\text{vac}} \phi_h^{\dagger}(z).
\end{equation}
Since the out-states are naturally located at $z = \infty$, the conjugated operator $\phi_h^{\dagger}(z)$ is obtained from $\phi_h(z)$ via the conformal transformation $z \mapsto 1/z$. Using the transformation law \eqref{eq:1dconphi}, we find that
\begin{equation} \label{eq:dagger1d}
    \phi_h^{\dagger}(z) = z^{-2h} \phi_h(1/z).
\end{equation}
The mode expansion of the conjugated field following from \eqref{eq:modephi1d} is
\begin{equation} \label{eq:daggerphih1d}
    \phi_h^{\dagger}(z) = \sum_{n \in \mathbb{Z}} z^{-n - h} \phi_n^{h\dagger} = \sum_{n \in \mathbb{Z}} z^{-n - h} \phi_{-n}^h,
\end{equation}
which leads to the identification
\begin{equation} \label{eq:phidagger1d}
    \phi_n^{h \dagger} = \phi_{-n}^h.
\end{equation}
The annihilation condition for $\bra{\text{vac}}$ is then
\begin{equation} \label{eq:<0|phi=01d}
    \bra{\text{vac}} \phi^h_n = 0 = \bra{\text{vac}} \phi^{h \dagger}_{-n}, \qquad (n < -h).
\end{equation}
This conjugation prescription ensures proper normalization of primary states. Using the standard two-point function of primary operators in one-dimensional CFT, we obtain
\begin{equation} \label{eq:<hh'>1d}
    \bra{h} h' \rangle = \delta_{h, h'},
\end{equation}
which follows from
\begin{equation}
    \langle \phi_h(\xi) \phi_{h'}(0) \rangle = \frac{\delta_{h,h'}}{\xi^{2h}}.
\end{equation}

It is important to note that these properties rely solely on the transformation $z \mapsto 1/z$, for which the Schwarzian derivative vanishes: $\{1/z; z\} = 0$. Therefore, the transformation law is the same for primary and quasi-primary operators under this map. As such, the results above can be repeated for quasi-primary fields. In particular, the BMS$_3$ currents $T(z)$ and $M(z)$ are quasi-primaries, and their mode operators satisfy
\begin{equation} \label{eq:conju1d}
    L_n^{\dagger} = L_{-n}, \qquad M_n^{\dagger} = M_{-n}.
\end{equation}
Moreover, the annihilation condition for the vacuum is
\begin{equation} \label{eq:LM0}
    L_n \ket{\text{vac}}= 0 = M_n \ket{\text{vac}}  \qquad \text{for} \quad n \geq -1.
\end{equation}
In particular, being annihilated by the subalgebra elements $L_{0 ,\pm 1}, M_{0, \pm 1}$ reflects the Poincar\'e invariance of the vacuum.

\subsection{Highest weight representations}
The definition of the vacuum state in \eqref{eq:LM0} and the primary operator \eqref{eq:HWS1}–\eqref{eq:Virhwrep} uniquely determines the annihilation conditions for the primary states:
\begin{equation} \label{eq:HW1d}
    L_n \ket{h} = M_n \ket{h} = 0 \quad (n \geq 1);
\end{equation}
and the eigen equations for $M_0$ and $L_0$:
\begin{equation} \label{eq:L0M0eigen}
L_0 \ket{h} = h \ket{h}, \qquad M_0 \ket{h} = \ket{h}.
\end{equation}
Remarkably, the eigenvalue of $M_0$ is fixed to $1$ due to the special realization of BMS$_3$ representation on $S^1$. Based on the group structure of BMS$_3$ \eqref{eq:BMS3stru}, since all coordinate transformations on $S^1$ belong to the superrotation sector, the supertranslations can only act as weight-shifting operators with the same effect on all primary operators. As a result, all primary states on $S^1$ share the same eigenvalue of $M_0$. This is distinct from the Carrollian or Galilean realization of BMS$_3$ symmetry, in which the eigenvalue of $M_0$ varies in different primary states. We will discuss this issue further in subsection \ref{sec:CompareBMS3}.

Consequently, the primary states are not annihilated by the negative modes $M_n, L_n (n < 0)$. The descendant states of a primary state $\ket{h}$ are obtained by acting on it with negative modes of the BMS$_3$ generators. A general descendant at level $N$ takes the form
\begin{equation} \label{eq:desc1d}
L_{-n_1} \cdots L_{-n_i} M_{-m_1} \cdots M_{-m_j} \ket{h}, \quad 1 \leq n_1 \leq \cdots \leq n_i,\ 1 \leq m_1 \leq \cdots \leq m_j,
\end{equation}
where the level of this descendant is defined by $N = n_1 + \cdots + n_i + m_1+ \cdots + m_j >0$. The primary state $\ket{h}$ and its descendants span an irreducible representation of BMS$_3$, or a module $\mathcal{V} (h, c_L, c_M)$. Descendants at each level $N$ can be grouped by their conformal weight $h + N$, as illustrated in Table~\ref{tab:Vmod1d}. Note that every descendant is an eigenstate of $L_0$, with eigenvalue (conformal weight) $h + N > h$. Thus, $h$ is the lowest conformal weight in the module. Therefore, $\mathcal{V} (h, c_L, c_M)$ is a HWR of BMS$_3$, 
labeled by the conformal weight $h$ of the primary state. We do not fix a priori which values of \(h\) in a given theory—those allowed values must be determined by the dynamics or consistency conditions (e.g., unitarity, boundary conditions) of that specific model. 

\begin{table}[htbp]
    \centering
    \begin{tabular}{cc}
         \hline
         $N$ & states $\{v_i^{(N)}\}$ \\
         \hline
         0 & \makecell[l]{$\ket{h}$} \\
         \hline
         1 & \makecell[l]{$L_{-1} \ket{h}; M_{-1} \ket{h}$} \\
         \hline
         2 & \makecell[l]{$L_{-2} \ket{h}, L_{-1}^2 \ket{h};$ \\ $M_{-2} \ket{h}, M_{-1}^2 \ket{h};$ \\ $L_{-1} M_{-1} \ket{h}$} \\
         \hline
         3 & \makecell[l]{$L_{-3} \ket{h}, L_{-1} L_{-2} \ket{h}, L_{-1}^3 \ket{h};$ \\ $M_{-3} \ket{h}, M_{-1} M_{-2} \ket{h}, M_{-1}^3 \ket{h};$ \\ $L_{-1} M_{-2} \ket{h}, L_{-2} M_{-1} \ket{h}, L_{-1}^2 M_{-1} \ket{h}, L_{-1} M_{-1}^2 \ket{h}$} \\
         \hline
         4 & \makecell[l]{$L_{-4} \ket{h}, L_{-1} L_{-3} \ket{h}, L_{-1}^2 L_{-2} \ket{h}, L_{-1}^4 \ket{h}, L_{-2}^2 \ket{h};$ \\ $M_{-4} \ket{h}, M_{-1} M_{-3} \ket{h}, M_{-1}^2 M_{-2} \ket{h}, M_{-1}^4 \ket{h}, M_{-2}^2 \ket{h};$ \vspace{5pt} \\ $L_{-1} M_{-3} \ket{h}, L_{-3} M_{-1} \ket{h}, L_{-1}^2 M_{-2} \ket{h}, L_{-1} L_{-2} M_{-1} \ket{h}, $ \\ $L_{-1} M_{-1} M_{-2} \ket{h}, L_{-2} M_{-1}^2 \ket{h}, L_{-1}^3 M_{-1} \ket{h}, L_{-1}^2 M_{-1}^2 \ket{h},$ \\ $L_{-1}^1 M_{-1}^3 \ket{h}, L_{-2} M_{-2} \ket{h}$} \\
         \hline
         $\vdots$ & \\
         \hline
    \end{tabular}
    \caption{The states in the highest weight BMS$_3$ module $\mathcal{V} (h, c_L, c_M)$. The level $N (\in \mathbb {Z} ^+)$ labels the descendants with conformal weight $h+N$.}
    \label{tab:Vmod1d}
\end{table}

Note that the order of BMS generators in \eqref{eq:desc1d} is chosen by convention. It is possible to construct states using negative modes $L_{n<0}, M_{n<0}$ like $\ket{h+N}_{ML} = M \cdots L \cdots \ket{h}$. Such states can be re-expressed using the algebra \eqref{eq:cBMS3A} as the linear combination of $\ket{h+N}_{LM} = L \cdots M \cdots \ket{h}$ with the same level $N$. In addition, the states that simultaneously include both negative and positive modes can also be non-vanishing when the positive modes $L_{n>0}, M_{n>0}$ are not attached to the primary state. Such states $L_{n} L_{-i} \cdots M_{-j} \cdots \ket{h} (0<n<N)$ can also be re-expressed by the negative-mode states at the level of $N-n$ (if $n>N$, then these states are vanishing). In this case, the central charge might appear. That is why each module is encoded by $(h, c_L, c_M)$. 

It is very important to point out that the highest weight vacuum is not the same as the $h = 0$ highest weight state introduced previously due to the difference of their annihilation conditions \eqref{eq:LM0} and \eqref{eq:HW1d}, \eqref{eq:L0M0eigen}. The descendants of the vacuum $\ket{\text{vac}}$ should be 
\begin{equation} \label{eq:descvac1d}
    L_{-n_1} \cdots L_{-n_i} M_{-m_1} \cdots M_{-m_j} \ket{\text{vac}}, \quad 2 \leq n_1 \leq \cdots \leq n_i,\ 2 \leq m_1 \leq \cdots \leq m_j
\end{equation}
which is slightly different from \eqref{eq:desc1d}: the descendants exclude the modes $L_{-1}$ and $M_{-1}$. Here, the order of BMS$_3$ generators is also chosen by convention. The vacuum state $\ket{\text{vac}}$ and its descendants also span an irreducible representation of BMS$_3$, or vacuum module $\mathcal{V}_{\text{vac}} (c_L, c_M)$ of BMS$_3$, which is different from $\mathcal{V} (h=0, c_L, c_M)$. Since the conformal weight of $\ket{\text{vac}}$ is the smallest one compared to the states in the vacuum module, which guarantees that the vacuum module is also a HWR.

\subsection{The inner product structure} \label{sec:Gram}

The inner products between the states from different modules vanish. Only the states at the same level can have non-zero overlap within a module. 
Here, we will present only examples of the inner product structure between states in the module $\mathcal{V} (h, c_L, c_M)$. The vacuum case can be similarly constructed.

To evaluate inner products among the states within Hilbert space, we will apply the Hermitian conjugation rules \eqref{eq:conju1d}, the commutation relations \eqref{eq:cBMS3A}, and the orthonormal condition \eqref{eq:<hh'>1d}. The overlaps between the states $\{v_i^{(N)}\}$ in the level $N$ define the Gram matrix as
\begin{equation}
    G_{ij}^{(N)} = \left\langle v_i^{(N)} \Big| v_j^{(N)} \right\rangle,
\end{equation}
where the order of the basis $v_i^{(N)}$ is given in the table \ref{tab:Vmod1d}. The Gram matrix encodes the inner product structure of the representation. 
For example, the Gram matrix at levels zero, one, and two is, respectively, given by
\begin{equation} \label{eq:G0}
    G^{(0)} = 1;
\end{equation}
\begin{equation}
G^{(1)} = 
\begin{pmatrix}
    2h& 2 \\ 2 & 0
\end{pmatrix};
\end{equation}
\begin{equation} \label{eq:G2}
    G^{(2)} = 
    \begin{pmatrix}
        \frac{c_L}{2} + 4h & 6h & \frac{c_M}{2} + 4 & 0 & 6 \\
        6h & 8 h^2 + 4h & 6 & 8 & 8h + 4\\
        \frac{c_M}{2} + 4 & 6 & 0 & 0 & 0 \\
        0 & 8 & 0 & 0 & 0 \\
        6 & 8h + 4 & 0 & 0 & 4
    \end{pmatrix}.
\end{equation}

Notice that the diagonal elements of the Gram matrix contain some zeros, because of the presence of zero-mode descendants. This necessitates a careful discussion of unitarity. Suppose $\ket{\chi}$ is a level-$N$ zero-norm descendant of the highest-weight state $\ket{h}$. If the zero-norm state $\ket{\chi}$ is null, i.e., it also satisfies the highest weight conditions \eqref{eq:HW1d} and \eqref{eq:L0M0eigen}, it should not be regarded as part of the module $\mathcal{V} (h, c_L, c_M)$; instead, it corresponds to a gauge redundancy associated with a highest-weight state $\ket{h+N}$ in a different module $\mathcal{V} (h+N, c_L, c_M)$. In this case, $\ket{h+N} + \ket{\chi}$ is also a weight-$(h+N)$ highest-weight state that cannot be distinguished with $\ket{h+N}$, as they possess identical inner product structure with all other states. Therefore, the zero-norm state like $\ket{\chi}$ should be modded out to obtain the physical Hilbert space. If all zero-norm descendants in the theory are of this type, then the resulting theory remains unitary. For the HWR of our construction, since the supertranslation sector is Abelian, all pure supertranslation descendants turn out to be zero-norm states. Unfortunately, they do not satisfy the highest-weight condition. Consequently, these zero-norm states cannot be interpreted as gauge redundancies and cannot be removed from the Hilbert space, implying that the HWRs constructed here are generically non-unitary.

\subsection{Characters of highest weight modules}

We will then calculate the characters of the HWRs. First, we compute the character of $\mathcal{V} (h, c_L, c_M)$. Since the vacuum module $\mathcal{V}_{\text{vac}} (c_L, c_M)$ differs from $\mathcal{V} (h=0, c_L, c_M)$, its character needs to be considered separately.

\subsubsection{Character of $\mathcal{V} (h, c_L, c_M)$}

The character of a certain highest-weight module $\mathcal{V} (h, c_L, c_M)$ of BMS$_3$ on $S^1$ is defined by
\begin{equation}
    \chi_{h, c_L, c_M} ({\bf{a}}, {\bf{b}}) = \text{tr}_{\mathcal{V} (h, c_L, c_M)} \left[ \text{exp} \left( \sum_{i \in \mathbb Z} a_i L_i + \sum_{m \in \mathbb Z} b_m M_m -\frac{c_L}{24} a_0 -\frac{c_M}{24} b_0 \right) \right],
\end{equation}
where ${\bf{a}}=\{a_0, \cdots ,a_i, \cdots \},\, {\bf{b}}=\{b_0, \cdots ,b_m, \cdots \}$ are some complex variables. It is also remarkable that the trace $\text{tr}_{\mathcal{V}}$ of an operator $\hat{A}$ is defined by the trace of its representation matrix in terms of the BMS$_3$ module $\mathcal{V}$
\begin{equation}
    \text{tr}_{\mathcal{V}} \hat{A} = \sum_{i \in \mathcal{V}} A_{ii}, \qquad \hat{A} \ket{i} = \sum_{j \in \mathcal{V}} A_{ij} \ket{j}.
\end{equation}
In general, the trace can also be defined using the inner product tr$(\hat A) = \sum_i \bra{i} \hat{A} \ket{i}$. However, this requires that the states in the module be orthonormal. As analyzed in the previous section regarding the inner product structure, there exist some zero-norm states (but not null) in the BMS$_3$ module that cannot be normalized. Therefore, we cannot use the inner product to define the trace in this context, as this will overlook some zero-norm states.

Note that the modes $M_m, L_n (m, n \neq 0)$ change the levels of the states, indicating that these modes have vanishing diagonal elements, and thus only $M_0, L_0$ contribute to the character. Conventionally, we choose $a_0 = a, \, b_0 = b$. Therefore, the character can be rewritten as 
\begin{equation} \label{eq:chih}
    \chi_{h, c_L, c_M} (a,b) = \text{tr}_{\mathcal{V} (h, c_L, c_M)} \left[ e^{- a (L_0 - \frac{c_L}{24})} e^{- b (M_0 - \frac{c_M}{24})} \right]. 
\end{equation}
The $L_0$ part is easy, as all states in the Hilbert space are the eigenstates of $L_0$. However, the $M_0$ part is relatively complex because the matrix of $M_0$ is not diagonal. Particularly, the supertranslation descendants $M_{-m_1} \cdots M_{-m_j} \ket{h}$ are eigenstates of $M_0$, as the supertranslation modes are abelian. However, the mixed descendants and Virasoro descendants are not the eigenstates of $M_0$, since the commutator of $M_0$ and the Virasoro generator will convert $L_n$ to $M_n$. In general, the action of $e^{-b M_0}$ can be expressed as
\begin{equation}
    \begin{split}
        & e^{-b M_0} L_{-n_1} \cdots L_{-n_i} M_{-m_1} \cdots M_{-m_j} \ket{h} \\
        & = e^{-b} L_{-n_1} \cdots L_{-n_i} M_{-m_1} \cdots M_{-m_j} \ket{h} + \cdots,
    \end{split}
\end{equation}
where ``$+ \cdots$'' includes the states that are different from $L_{-n_1} \cdots L_{-n_i} M_{-m_1} \cdots M_{-m_j} \ket{h}$, which contribute to the non-diagonal terms in the matrix of $e^{-b M_0}$. Hence, the diagonal elements of the matrix $e^{-b M_0}$ are all $e^{-b}$, independent of the conformal weights of the states in the module. Consequently, the character can be re-expressed as
\begin{equation}
    \chi_{h, c_L, c_M} (a,b) = e^{- b (1 - \frac{c_M}{24})} e^{ a \frac{c_L}{24}} \sum_{h_{\mathcal{V}} \geq h} e^{- a h_{\mathcal{V}}} \text{dim}_{h_{\mathcal{V}}},
\end{equation}
where dim$_{h_{\mathcal{V}}}$ denotes the number of the weight-$h_{\mathcal{V}}$ states within the module $\mathcal{V} (h, c_L, c_M)$.

The general states in \eqref{eq:desc1d} that share the same conformal weight $h_{\mathcal{V}} = h + n_L + n_M$ have $p (n_L) \times p(n_M)$ choices. Here, $n_L = n_1 + \cdots + n_i$ and $n_M = m_1 + \cdots + m_j$ are fixed but $n_1, \cdots ,n_i; ~m_1, \cdots, m_j$ are variables. Moreover, $p (n)$ counts the distinct partitions of the integer $n$, which can be generated from the series expansion of the Euler function:
\begin{equation} \label{eq:Euler}
    \frac{1}{\varphi (x)} = \prod^\infty_{n = 1} \frac{1}{1-x^n} = \sum_{n = 0}^\infty p (n) x^n.
\end{equation}
Then, the character can be derived as
\begin{equation} \label{eq:character1d}
    \begin{split}
        \chi_{h, c_L, c_M} (a, b) 
        & = e^{- b (1 - \frac{c_M}{24})} e^{a \frac{c_L}{24}} \sum_{n_L, n_M = 0}^\infty p(n_L) p(n_M) e^{- a (h + n_L + n_M)} \\
        & = 
        \frac{e^{- b (1 - \frac{c_M}{24})} e^{- a (h - \frac{c_L}{24})}}{\prod^\infty_{n = 1} (1 - e^{-n a})^2}.
    \end{split}
\end{equation}

\subsubsection{Character of $\mathcal{V}_{\text{vac}} (c_L, c_M)$ and the gravitational partition function} \label{sec:3Dbulkrep}

It is significant to compute the character of the highest-weight vacuum module, as it has clear connections to the partition functions of the bulk gravity theory in AdS$_3$ case \cite{Witten:2007kt,Maloney:2007ud,Giombi:2008vd,David:2009xg}. Based on the difference between the vacuum descendants \eqref{eq:descvac1d} and the $h = 0$ case in \eqref{eq:desc1d}, the character of the vacuum module can be derived by excluding all contributions that contain $L_{-1}$ or $M_{-1}$ in $\chi_{0,c_L, c_M} (a,b)$, the $h = 0$ case in \eqref{eq:character1d}, which causes a prefactor $(1- e^{- a})^2$. 
Additionally, the contribution of $M_0$ in $\chi_{0,c_L, c_M}$ should also be removed when considering the character of the vacuum module, since $M_0$ annihilates the vacuum state \eqref{eq:LM0}, unlike its action on the highest weight state \eqref{eq:L0M0eigen}. This can be achieved by multiplying by an extra prefactor $e^b$. Eventually, one obtains the vacuum character
\begin{equation} \label{eq:vacchi1d}
    \begin{split}
        \chi_{\text{vac};\, c_L,c_M}&  (a,b) = \text{tr}_{\mathcal{V}_{\text{vac}}} \left[ e^{- a (L_0 - \frac{c_L}{24})} e^{- b (M_0 - \frac{c_M}{24})} \right] \\
        = (1 - &e^{- a})^2e^b \chi_{0, c_L, c_M} (a,b) = \frac{e^{ a \frac{c_L}{24} + b \frac{c_M}{24}}}{\prod^\infty_{n = 2} (1- e^{- n a})^2}.
    \end{split}
\end{equation}

Now, we compare the character from the vacuum module in the HWR with the one from the vacuum module in the induced representation \cite{Oblak:2015sea}. The induced vacuum $\ket{\text{vac}}_I$ is defined as \cite{Campoleoni:2016vsh}
\begin{equation}
    L_0 \ket{\text{vac}}_I = L_{\pm 1} \ket{\text{vac}}_I =0; \qquad M_n \ket{\text{vac}}_I = 0 \quad (\forall n \in \mathbb Z).
\end{equation}
Its descendants are then given by
\begin{equation} \label{eq:inducedvacuum}
    L_{-n_1} \cdots L_{-n_i} L_{m_1} \cdots L_{m_j} \ket{\text{vac}}_I, \quad 2 \leq n_1 \leq \cdots \leq n_i,\ 2 \leq m_1 \leq \cdots \leq m_j.
\end{equation}
Therefore, there is a bijection between the descendants of the induced vacuum and those of the highest weight vacuum \eqref{eq:descvac1d}: the supertranslation generators $M_{-m} (m>1)$ in \eqref{eq:descvac1d} are replaced by $L_m (m>1)$ in the induced vacuum module. However, the inner product between states in the induced module is all orthonormal and positive definite, indicating that the induced representations are unitary \cite{Barnich:2014kra,Barnich:2015uva, Oblak:2015sea}, which are totally different from the HWRs of our construction. The discrepancy between the two types of representations is also reflected in their characters. For example, the character $\chi^{(I)}_{\text{vac};\, c_L,c_M}$ of the vacuum module in the induced representation is obtained from \eqref{eq:inducedvacuum} as
\begin{equation} \label{eq:HWvac=in}
    \chi^{(I)}_{\text{vac};\, c_L,c_M} (a,b) = \frac{e^{ a \frac{c_L}{24} + b \frac{c_M}{24}}}{\prod^\infty_{n = 2} (1- e^{- n a}) (1- e^{n a})}.
\end{equation}
The two factors $\prod^\infty_{n = 2} (1-e^{n a})^{-1}$ and $\prod^\infty_{n = 2} (1-e^{- n a})^{-1}$ in the induced vacuum character above arise from Virasoro modes with positive and negative indices, respectively. In contrast, the BMS generators $M_{-m}, L_{-m} (m >1)$ that generate the descendants of the highest-weight vacuum all have negative indices, resulting in two identical factors $(1-e^{- n a})^{-1}$ in \eqref{eq:vacchi1d}. More explicitly, one can show that the two vacuum characters differ up to a prefactor
\begin{equation} \label{eq:HWvac=indu}
    \chi^{(I)}_{\text{vac};\, c_L,c_M} (a,b)= e^{-\sum_{n=2}^\infty (a n + i \pi)} \chi_{\text{vac};\, c_L,c_M} (a,b),
\end{equation}
which explicitly showcases the difference between the induced representations and HWRs in our constructions.

The partition function of 3D Einstein's gravity theory, fluctuating near the standard Minkowski metric, can be perturbatively expanded by the power of the Planck constant $\hbar$. The result that is accurate to one loop is derived in \cite{Barnich:2015mui} by the Euclidean path integral:
\begin{equation} \label{eq:1loopgrav}
    Z_{\text{grav}}^{(E)} = \frac{e^{\frac{\beta}{8 G \hbar}}}{\prod^\infty_{n = 2} |1- e^{i n \theta}|^2},
\end{equation}
where $G$ is Newton's constant, and the metric of the vacuum $\mathbb R^3/ \gamma$ is
\begin{equation} 
    \td s^2 = \td y^2 + \td \rho^2 + \rho^2 \td \varphi^2, \quad \gamma: (y, \varphi) \sim (y+\beta, \varphi + \theta),
\end{equation}
where $\theta, \beta$ are regarded as the angular potential and the inverse of temperature, which are the periods of $\varphi$ and the Euclidean time $y$, respectively. 

On the other hand, the periodicity of spacetime allows the partition function to be expressed via the path integral as
\begin{equation}
    Z_{\text{grav}}^{(E)} = \int \td^3 x \bra{x} e^{-i \theta (L_0 - \frac{c_L}{24}) - \beta (M_0 - \frac{c_M}{24})} \ket{x} = \text{tr} e^{-i \theta (L_0 - \frac{c_L}{24}) - \beta (M_0 - \frac{c_M}{24})},
\end{equation}
where $e^{-i \theta (L_0 - \frac{c_L}{24}) - \beta (M_0 - \frac{c_M}{24})}$ denotes the spacetime translation operator. This expression resembles the definition of the characters \eqref{eq:chih} and \eqref{eq:vacchi1d} with $a = i \theta (0 \leq \theta < 2 \pi), \, b = \beta \in \mathbb R$. Note that 3D gravity has no propagating degree of freedom; its partition function near vacuum can only correspond to the character of the vacuum modules of the possible dual theory. To illustrate this more explicitly, we rewrite the character of the highest-weight vacuum module \eqref{eq:vacchi1d} as 
\begin{equation}
    \chi_{\text{vac};\, c_L,c_M} (i \theta, \beta) = \frac{e^{i\theta \frac{c_L}{24} + i \sum_{n = 2}^\infty (n \theta + \pi) + \beta \frac{c_M}{24}}}{\prod^\infty_{n = 2} |1- e^{i n \theta}|^2}.
\end{equation}
Then, the above highest-weight vacuum character coincides with the gravitational partition function \eqref{eq:1loopgrav} as
\begin{equation} \label{eq:HWcha=Z}
    \chi_{\text{vac};\, c_L = 26,c_M = 3} (i \theta, \beta) = i Z_{\text{grav}}^{(E)}\,,
\end{equation}
where we have chosen the natural units to set $G = \hbar = 1$, and we used the analytic continuation of Riemann $\zeta$-function $\zeta (s) = \sum_{n=1}^\infty \frac{1}{n^s}$ to regularize the divergent summations $\sum_{n = 2}^\infty n \rightarrow \zeta (-1) - 1 = -\frac{13}{12}$ and $\sum_{n = 2}^\infty 1 \rightarrow \zeta (0) - 1 = -\frac{3}{2}$. This concrete relationship provides strong evidence for flat holography in three dimensions, i.e., the duality between the 3D bulk gravity theory and the CFT with the central charge $c_M = 3,~ c_L= 26$ on the codimension-two boundary $S^1$.\footnote{Actually, we study the HWR of the central extended BMS$_3$ algebra with arbitrary central change in this section, as the representation theory itself does not determine the central charges. Different central charges can lead to distinct field theories on $S^1$. The resulting character is valid for any choice of the central charge and different values of $c_L$ only modify the character by a phase factor. In particular, choosing $c_L=26$ yields a constant, $\theta$-independent phase factor ``$i$'', which, together with $c_M = 3$,  ensures that the character coincides with the gravitational partition function up to this phase factor.} This relation also offers valuable insights and motivation for the discussion of BMS$_4$ in the following section.

In the seminal works \cite{Barnich:2015mui,Oblak:2015sea}, it is revealed that the partition function \eqref{eq:1loopgrav} matches the character of induced vacuum module \eqref{eq:HWvac=in} with the central charges $c_M = 3, c_L = 0$, and $a = i \theta, \, b = \beta$:
\begin{equation}
    \chi_{\text{vac};\, c_L = 0, c_M = 3}^{(I)} (i \theta, \beta) = Z_{\text{grav}}^{(E)}\,,
\end{equation}
which differs from the case of the highest-weight vacuum, where $c_L=26$. This difference is anticipated, since the induced representation and HWR give rise to different vacuum representations as discussed previously.

\subsection{Comparison with 2D Carrollian} \label{sec:CompareBMS3}

In this subsection, we compare the BMS$_3$-invariant field theories on $S^1$ with the one on a 2D Carrollian manifold. The latter defines the two-dimensional Carrollian CFT (CarCFT$_2$). 
The structure of a 2D Carrollian manifold is $\mathbb R_u \times S^1$, where $u$ is a null temporal direction. 
In a 2D Carrollian manifold, the supertranslation action can manifest directly through a local translation in $u$.

The BMS$_3$ transformation for the primary fields on the 2D Carrollian manifold is given by \cite{Hao:2021urq, Hao:2022xhq}
\begin{equation} \label{eq:Carrolliantrans}
    [L_n, \bar \phi_{\Delta, \boldsymbol{\xi}} (u,z)] = l_n \bar \phi_{\Delta, \boldsymbol{\xi}} (u,z) \,, \qquad [M_n, \bar \phi_{\Delta, \boldsymbol{\xi}} (u,z)] = m_n \bar \phi_{\Delta, \boldsymbol{\xi}} (u,z)\,,
\end{equation}
with
\begin{equation} \label{eq:HWRCar2}
    \begin{split}
        l_n & = - z^{n+1} \partial_z - (n+1) z^n u \partial_u - (n+1) \left(z^n \Delta + n z^{n-1} u \boldsymbol{\xi} \right), \\
        m_n & = z^{n+1} \partial_u + (n+1) z^n \boldsymbol{\xi}.
    \end{split}
\end{equation}
Here, the primary operators $\bar \phi_{\Delta, \boldsymbol{\xi}} (u,z)$ represent the fields in 2D Carrollian manifold, and they are characterized by the scaling dimension $\Delta$ and the boost charge $\xi$, which is the eigenvalue of the non-diagonalizable matrix $\boldsymbol{\xi}$, and are generally multiplets. The non-diagonalizable matrix $\boldsymbol{\xi}$ can be chosen as the Jordan form
\begin{equation}
    \boldsymbol{\xi} = 
    \begin{pmatrix}
        \xi \\
        1 & \xi \\
        & \ddots & \ddots \\
        & & 1 & \xi
    \end{pmatrix}_{r \times r},
\end{equation}
where $r$ denotes the number of operators in a multiplet. Correspondingly, the primary state should be denoted as $\ket{\Delta, \boldsymbol{\xi}}$. Following the same approach as the previous subsections, one can derive that
\begin{equation} \label{eq:LMphiCar2}
    \begin{split}
        & L_n \ket{\Delta, \boldsymbol{\xi}} = M_n \ket{\Delta, \boldsymbol{\xi}} = 0 \qquad ~~~~~~~~(n \geq 1), \\
        & L_0 \ket{\Delta, \boldsymbol{\xi}} = \Delta \ket{\Delta, \boldsymbol{\xi}}, \quad M_0 \ket{\Delta, \boldsymbol{\xi}} = \boldsymbol{\xi} \ket{\Delta, \boldsymbol{\xi}},
    \end{split}
\end{equation}
where the highest weight state $\ket{\Delta, \boldsymbol{\xi}}$ corresponds to the Carrollian primary field $\bar \phi_{\Delta, \boldsymbol{\xi}} (u,z)$ through the state-operator correspondence of the 2D Carrollian CFT. The multiplet structure prevents $M_0$ and $L_0$ from having a common eigenbasis if $r > 1$, as $\boldsymbol{\xi}$ cannot be diagonalized if the highest-weight multiplet has more than one component. Then, an HWR of BMS$_3$  realized on the 2D Carrollian manifold is composed of a multiplet primary state $\ket{\Delta, \boldsymbol{\xi}}$ and its descendants. In particular, for $r = 1$, i.e., the singlets, the primary states become the common eigenstates of $M_0, L_0$ with the eigenvalue $\xi, \Delta$ varying between different primary states. This differs from the HWR of BMS$_3$ realized on $S^1$, where the eigenvalue of $M_0$ is identical for all primary states. This is because supertranslations in the Carrollian case generate translations along the $u$-direction on the Carrollian manifold, requiring an additional quantum number, i.e., the boost charge $\xi$, to characterize the supertranslation and to distinguish different primary states.

Comparing \eqref{eq:Carrolliantrans} and \eqref{eq:HWRCar2} with \eqref{eq:HW1d} and \eqref{eq:L0M0eigen}, the HWR of BMS$_3$ realized on $S^1$ is isomorphic to the HWR constructed from singlet primary states with $\xi = 1, \Delta = 0$ on a 2D Carrollian manifold. The map between them is the modified Mellin transform
\begin{equation} \label{eq:dictCCFT1}
    \int^{\infty}_{- \infty} \td u u^{-h} \bar \phi_{0,1} (u, z) = \phi_h (z),
\end{equation}
which connects the singlet Carrollian field $\bar \phi_{0,1} (u, z)$ to the primary field $\phi_h (z)$ on $S^1$, where this transform translates $u, u \p_u$ to $e^{\p_h}, h$, respectively. Therefore, the change of the conformal weight $h$ in CCFT$_1$ corresponds to a pure $u$-transform in the Carrollian manifold. The dictionary \eqref{eq:dictCCFT1} agrees with the discussions in \cite{Donnay:2022aba, Donnay:2022wvx, Bagchi:2022emh} \footnote{These references built the correspondence between 3D Carrollian and 2D celestial. Fortunately, in our case, although the spatial dimension is reduced by one, this dictionary remains valid, as the transform for time is independent of the spatial dimensions.}.

\section{Highest weight representation of BMS$_4$ realized on $S^2$} \label{sec:BMS4}

The BMS$_4$ group is a semidirect product structure of the superrotations and the supertranslations
\begin{equation} \label{eq:BMS4A}
    \mathrm{BMS}_4 = (\text{Diff} (S^1) \times \text{Diff} (S^1)) \ltimes C^{\infty} (S^2).
\end{equation}
where the superrotation sector $\text{Diff} (S^1) \times \text{Diff} (S^1)$ is the local conformal group of $S^2$, including all spacetime conformal transformations of $S^2$, and the supertranslation sector $C^\infty (S^2)$ is characterized by arbitrary functions on the two-dimensional celestial sphere $S^2$. Consequently, the actions of supertranslation on $S^2$ can only shift the conformal weight, similar to the BMS$_3$ case. 

In this section, we will construct the HWR of the extended BMS$_4$ algebra \cite{Barnich:2010eb, Barnich:2011mi, Barnich:2017ubf}
\begin{equation} \label{eq:BMSA}
    \begin{split}
        & [L_n, L_m] = (n-m) L_{m+n}, \quad [\Bar{L}_n, \Bar{L}_m] = (n-m) \Bar{L}_{m+n}, \\
        & [L_n, M_{r,s}] = \left( \frac{n}{2} - r \right) M_{r + n, s}, \quad [\Bar{L}_n, M_{r,s}] = \left( \frac{n}{2} - s \right) M_{r, s + n}, \\
        & [L_n, \Bar{L}_m] = 0, \quad [M_{r,s}, M_{t, u}] = 0, \quad r, s, t, u \in \mathbb{Z} + \frac{1}{2},
    \end{split}
\end{equation}
through the BMS$_4$-invariant field theory on $S^2$, following the same approach of section \ref{sec:BMS3}. Here, $(L_n, \Bar{L}_n)$ and $M_{r, s}$ are the generators of superrotation and supertranslation respectively. The BMS$_4$ algebra in \eqref{eq:BMSA} has no central charge, and the superrotation aspect of this algebra $L_n, \bar L_n$ consists of two sets of Witt algebras, also known as the Virasoro algebras with vanishing central charges. Therefore, the BMS$_4$-invariant field theory on $S^2$ must be a two-dimensional CFT without central charge. The additional symmetries, namely the supertranslations, shift the conformal weight. We will compute the characters of the highest-weight modules, which could provide valuable insights into the four-dimensional gravitational partition function.

\subsection{BMS$_4$ transformations and OPEs in field theory on $S^2$} \label{sec:HW2d}

The operator that contains all information about the Virasoro sectors is the stress tensor, with the following mode expansion
\begin{equation} \label{eq:Texpand}
    \begin{split}
        & T_{zz} = \sum_{n \in \mathbb Z} z^{-n-2} L_n, \quad T_{\Bar{z} \Bar{z}} = \sum_{n \in \mathbb Z} \bar z^{-n-2} \bar L_n; \\
        & L_n = \oint_0 \frac{\td z}{2 \pi i} z^{n+1} T_{zz}, \quad \Bar{L}_n = \oint_0 \frac{\td \bar z}{2 \pi i} \bar z^{n+1} T_{\bar z \bar z}.
    \end{split}
\end{equation}
The operator that encapsulates all supertranslation modes can be constructed as \cite{Barnich:2017ubf} 
\begin{equation} \label{eq:Mscalar}
    \begin{split}
        & M (z, \Bar{z}) = \sum_{m,n \in \mathbb Z} z^{-m -1} \Bar{z}^{-n -1} M_{m - \frac{1}{2},n - \frac{1}{2}}, \\
        & M_{m - \frac{1}{2},n - \frac{1}{2}} = \oint_0 \frac{\td z}{2 \pi i} z^n \oint_0 \frac{\td \Bar{z}}{2 \pi i} \Bar{z}^m M (z, \Bar{z}),
    \end{split}
\end{equation}
which is a scalar operator with conformal weight $(\frac{3}{2}, \frac{3}{2})$. Note that the mode expansion $M(z, \bar z)$ incorporates the fractional parts of $h, \bar h$ into the coefficients to avoid the emergence of multivalued functions, which differs slightly from the conventional form below in \eqref{eq:modephiS2}. The OPEs between the stress tensor and $M (z, \bar z)$ can be derived from the BMS$_4$ algebra by counting the residue
\begin{equation} \label{eq:TM}
    \begin{split}
        & T_{zz} T_{ww} \sim \frac{2 T_{ww}}{(z - w)^2} + \frac{\p_w T_{ww}}{z - w}, \quad T_{\Bar{z} \Bar{z}} T_{\Bar{w} \Bar{w}} \sim c.c., \\
        & T_{zz} M (w, \Bar{w}) \sim \frac{3/2}{(z - w)^2} M (w, \Bar{w}) + \frac{\p_w M (w, \Bar{w})}{z - w}, \\
        & T_{\Bar{z} \Bar{z}} M (w, \Bar{w}) \sim \frac{3/2}{(\Bar{z} - \Bar{w})^2} M (w, \Bar{w}) + \frac{\p_{\Bar{w}} M (w, \Bar{w})}{\Bar{z} - \Bar{w}}, \\
        & T_{zz} T_{\bar w \bar w} \sim 0, \qquad M (z, \bar z) M(w, \bar w) \sim 0,
    \end{split}
\end{equation}
which is consistent with the results from the soft limit of amplitude \cite{Pate:2019lpp, Fotopoulos:2019vac}. Therefore, the scalar operator $M (z, \bar z)$ and the stress tensor components are all primary operators because there is no central charge.

Under the BMS$_4$ transformation, the primary field $\phi_{h, \Bar{h}} (z, \Bar{z})$ with conformal weights $(h, \bar h)$ is transformed as follows  \cite{Fotopoulos:2019vac}
\begin{equation} \label{eq:basis}
    \begin{split}
        & [L_n, \phi_{h, \Bar{h}} (z, \Bar{z})] = \left[ z^{n+1} \p_z + (n + 1) z^n h \right] \phi_{h, \Bar{h}} (z, \Bar{z}), \\
        & \left[ M_{m - \frac{1}{2}, n - \frac{1}{2}}, \phi_{h, \Bar{h}} (z, \Bar{z}) \right] = z^m \Bar{z}^n e^{\frac{\p_h + \p_{\Bar{h}}}{2}} \phi_{h, \Bar{h}} (z, \Bar{z}).
    \end{split}
\end{equation}
The OPE between the stress tensor and $\phi_{h, \Bar{h}}$ can be derived from the above commutators by counting the residues
\begin{equation} \label{eq:Tphi}
    \begin{split}
        & T_{zz} \phi_{h, \Bar{h}} (w, \Bar{w}) \sim \frac{h}{(z - w)^2} \phi_{h, \Bar{h}} (w, \Bar{w}) + \frac{1}{z - w} \p_w \phi_{h, \Bar{h}} (w, \Bar{w}), \\
        & T_{\Bar{z} \Bar{z}} \phi_{h, \Bar{h}} (w, \Bar{w}) \sim \frac{\Bar{h}}{(\Bar{z} - \Bar{w})^2} \phi_{h, \Bar{h}} (w, \Bar{w}) + \frac{1}{\bar z - \bar w} \p_{\Bar{w}} \phi_{h, \Bar{h}} (w, \Bar{w}).
    \end{split}
\end{equation} 
Similarly, the $M \phi_{h, \Bar{h}}$ OPE can be derived as
\begin{equation} \label{eq:Mphi2d}
    M (z, \Bar{z}) \phi_{h, \Bar{h}} (w, \Bar{w}) \sim \frac{\phi^{M (1)}_{h + \frac{1}{2}, \Bar{h} + \frac{1}{2}} (w, \Bar{w})}{(z - w) (\Bar{z} - \Bar{w})}, \quad \phi^{M (1)}_{h + \frac{1}{2}, \Bar{h} + \frac{1}{2}} = e^{\frac{\p_h + \p_{\Bar{h}}}{2}} \phi_{h, \Bar{h}}.
\end{equation}
The OPEs can be equivalently expressed in the form of Ward identities
\begin{equation} \label{eq:WardTMS2}
    \begin{split}
        \<T_{zz} X_n\> & = \sum_{k = 1}^n \left[ \frac{h_k}{(z - z_k)^2} + \frac{\p_{z_k}}{z - z_k} \right] \<X_n\>, \\
        \<T_{\bar z \bar z} X_n\> & = \sum_{k = 1}^n \left[ \frac{h_k}{(\bar z - \bar z_k)^2} + \frac{\p_{\bar z_k}}{\bar z - \bar z_k} \right] \<X_n\>, \\
        \<M (z, \bar z) X_n\> & = \sum_{k = 1}^n \frac{1}{|z - z_k|^2} e^{\frac{\p_{h_k} + \p_{\bar h_k}}{2}} \<X_n\>,
    \end{split}
\end{equation}
where $X_n = \prod^n_{i = 1} \phi_{h_i, \bar h_i} (z_i, \bar z_i)$. Here, $\phi^{M (1)}_{h + \frac{1}{2}, \Bar{h} + \frac{1}{2}}$ is no longer a conformal primary operator. The mode expansion of primary operator $\phi_{h, \bar h}$ is 
\begin{equation}
    \phi_{h, \Bar{h}} (z, \Bar{z}) = \sum_{n,m \in \mathbb Z} z^{-n -h} \Bar{z}^{-m -\Bar{h}} \phi_{h, \Bar{h}}^{n,m},
\end{equation}
while the mode expansion of $\phi^{M (1)}_{h + \frac{1}{2}, \bar h + \frac{1}{2}}$ is
\begin{equation} \label{eq:phi'}
    \begin{split}
        \phi^{M (1)}_{h + \frac{1}{2}, \Bar{h} + \frac{1}{2}} (z, \Bar{z}) & = \sum_{n,m \in \mathbb Z} z^{-n -h -\frac{1}{2}} \Bar{z}^{-m - \Bar{h} -\frac{1}{2}} \phi_{h, \Bar{h}}^{n,m} \\ 
        & = |z|^{-1} \phi_{h, \Bar{h}} (z, \Bar{z}) .
    \end{split}
\end{equation}
Therefore, $\phi^{M (1)}$ is not a primary operator. From the above relation and the transformations in \eqref{eq:basis}, one can verify that the action of the supertranslation on the highest-weight states is null, which is significantly different from the BMS$_3$ case and will have an essential effect on the results of the character computation, as will be detailed in the following pages.

The charges generating BMS$_4$ transformations on $S^2$ are defined as
\begin{equation} \label{eq:BMS4Q2D}
    \begin{split}
        Q_\epsilon & = \oint \frac{\td z^A}{2 \pi i} T_{AB} \epsilon^B = \sum_{n \in \mathbb Z} (\epsilon_n L_n + \bar \epsilon_n \bar L_n), \\
        Q_f & = \oint \frac{\td z}{2 \pi i} \oint \frac{\td \bar z}{2 \pi i} M (\vec z) f (\vec z) = \sum_{n, m \in \mathbb Z} f_{n, m} M_{n - \frac12, m - \frac12},
    \end{split}
\end{equation}
where the coordinates $z$ and $\bar z$ are treated as independent variables when dealing with the contour integrals $\oint \frac{\td z}{2 \pi i}, \oint \frac{\td \bar z}{2 \pi i}$. We also used the shorthand $z^A = \vec z = (z, \bar z)$, and $\epsilon^A = (\epsilon (z), \bar \epsilon (\bar z))$ for notational simplicity. We have used the mode expansions
\begin{equation}
    \epsilon (z) = \sum_{n \in \mathbb Z} \epsilon_n z^{n+1}, \quad f (\vec z) = \sum_{n, m \in \mathbb Z} f_{n, m} z^n \bar z^m,
\end{equation}
and $\bar \epsilon (\bar z)$ can be expanded in a similar way. The infinitesimal BMS$_4$ transformation of the primary operators can be computed as
\begin{equation} \label{eq:[Q,phi]2D}
    \begin{split}
        [Q_\epsilon, \phi_{h, \bar h} (\vec z)] & = \Big[ h \p_z \epsilon (z) + \epsilon (z) \p_z + \bar h \p_{\bar z} \bar \epsilon (\bar z) + \bar \epsilon (\bar z) \p_{\bar z} \Big] \phi_{h, \bar h} (\vec z), \\
        [Q_f, \phi_{h, \bar h} (\vec z)] & = f (\vec z) \phi^M_{h + \frac{1}{2}, \bar h + \frac{1}{2}} (\vec z).
    \end{split}
\end{equation}
This is a rewriting of \eqref{eq:basis} from the charge perspective.

\subsection{Asymptotic states}

The primary state $\ket{h, \bar h}$ can be defined from the corresponding primary operator $\phi_{h, \bar h} (\vec z)$ through the state-operator correspondence
\begin{equation}
    \ket{h, \bar h} = \lim_{z, \bar z \rightarrow 0} \phi_{h, \bar h} (z, \bar z) \ket{\text{vac}}.
\end{equation}
The Hermitian conjugation of the primary operator can be defined by imposing a transformation $z \rightarrow \frac{1}{\bar z}$ to $\phi_{h, \bar h} (z, \bar z)$ as 
\begin{equation} \label{eq:phidaggerdef}
    \left[ \phi_{h, \bar h} (z, \bar z) \right]^{\dagger} = \bar z^{-2h} z^{-2 \bar h} \phi_{h, \bar h} (1/\bar z, 1/z),
\end{equation}
and thus its corresponding state is
\begin{equation}
    \bra{h, \bar h} = \ket{h, \bar h}^{\dagger} = \lim_{z \rightarrow 0} \bra{\text{vac}} \left[ \phi_{h, \bar h} (z, \bar z) \right]^{\dagger}.
\end{equation}
Based on the result of the two-point function for 2D CFT, the above definition guarantees the inner product between in and out states to be normalized
\begin{equation}
    \bra{h, \bar h} h', \bar h'\> = \delta_{h, h'} \delta_{\bar h, \bar h'}.
\end{equation}
Inserting \eqref{eq:phidaggerdef} into the mode expansions
\begin{equation} \label{eq:modephiS2}
\phi_{h, \bar h} (z, \bar z) = \sum_{n, m \in \mathbb Z} \phi_{m, n} z^{-m - h} \bar z^{-n -\bar h}, \quad \left[ \phi_{h, \bar h} (z, \bar z) \right]^{\dagger} = \sum_{n, m \in \mathbb Z} \phi_{m, n}^{\dagger} \bar z^{-m - h} z^{-n -\bar h},
\end{equation}
one obtains
\begin{equation} \label{eq:phidagger}
    \phi_{m,n}^{\dagger} = \phi_{-m,-n}.
\end{equation}
To ensure that the states $\ket{h, \bar h}$ are well-defined, the divergent terms in the mode expansions of the primary operator $\phi_{h, \bar h}$ as the coordinate approaches the origin must be annihilated by the vacuum. 
Therefore, any terms with negative powers in either of the variables $z$ or $\bar z$ must vanish when acting on the vacuum state, which leads to
\begin{equation}
    \phi_{m, n} \ket{\text{vac}} = 0, \quad m > -h, ~ \text{or} ~ n>- \bar h.
\end{equation}

Given that $M(\vec z), T_{AB}$ defined in \eqref{eq:Mscalar} and \eqref{eq:Texpand} are primary operators, the BMS$_4$ generators should satisfy
\begin{equation} \label{eq:MLdagger}
    M_{m - \frac{1}{2}, n - \frac{1}{2}}^{\dagger} = M_{- m + \frac{1}{2}, -n + \frac{1}{2}}, \qquad L_n^\dagger = L_{-n};
\end{equation}
and
\begin{equation} \label{eq:ML0=0}
    \begin{split}
        & M_{m-\frac{1}{2}, n-\frac{1}{2}} \ket{\text{vac}} = 0 \quad (m \geq 0 ~~ \text{or} ~~ n \geq 0), \\
        & L_n \ket{\text{vac}} = 0 = \bar L_n \ket{\text{vac}} \quad (n \geq -1);
    \end{split}
\end{equation}
or its Hermitian conjugate
\begin{equation} \label{eq:ML0=0+}
    \begin{split}
        & \bra{\text{vac}} M_{m-\frac{1}{2}, n-\frac{1}{2}} = 0 \quad (m \leq 1 ~~ \text{or} ~~ n \leq 1), \\
        & \bra{\text{vac}} L_n= 0 = \bra{\text{vac}} \bar L_n  \quad (n \leq 1).
    \end{split}
\end{equation}
In particular, the annihilations include the Poincar\'e transformations of the vacuum \cite{Stieberger:2018onx, Fotopoulos:2019vac}
\begin{equation}
    \begin{split}
        & M_{-\frac{1}{2}, -\frac{1}{2}} \ket{\text{vac}} = M_{\frac{1}{2}, -\frac{1}{2}} \ket{\text{vac}} = M_{-\frac{1}{2}, \frac{1}{2}} \ket{\text{vac}} = M_{\frac{1}{2}, \frac{1}{2}} \ket{\text{vac}} = 0, \\
        & L_{-1} \ket{\text{vac}} = L_0 \ket{\text{vac}} = L_1 \ket{\text{vac}} = \bar L_{-1} \ket{\text{vac}} = \bar L_0 \ket{\text{vac}} = \bar L_1 \ket{\text{vac}} = 0.
    \end{split}
\end{equation}

\subsection{Highest weight representation}

From the realization of BMS$_4$ on $S^2$, a primary state in the Hilbert space can now be characterized by its conformal weights $(h, \bar h)$. The highest-weight condition can be derived from \eqref{eq:basis} and \eqref{eq:ML0=0} as 
\begin{equation} \label{eq:LMh0}
    \begin{split}
        & L_n \ket{h, \bar h} = \bar L_n \ket{h, \bar h} = 0~~~~~~~~~~~~~~~~ (n > 0), \\
        & M_{m+\frac{1}{2}, n+\frac{1}{2}} \ket{h, \bar h} = 0 \qquad (m \geq 0 ~~ \text{or} ~~ n \geq 0), \\
        & L_0 \ket{h, \bar h} = h \ket{h, \bar h}, \qquad \bar L_0 \ket{h, \bar h} = \bar h \ket{h, \bar h};
    \end{split}
\end{equation}
or its Hermitian conjugate
\begin{equation} \label{eq:LMh0+}
    \begin{split}
        & \bra{h, \bar h} L_n = \bra{h, \bar h} \bar L_n = 0~~~~~~~~~~~~~~~~ (n < 0), \\
        & \bra{h, \bar h} M_{m+\frac{1}{2}, n+\frac{1}{2}} = 0 \qquad (m \leq -1 ~~ \text{or} ~~ n \leq -1), \\
        & \bra{h, \bar h} L_0 = \bra{h, \bar h} h, \qquad \bra{h, \bar h} \bar L_0 = \bra{h, \bar h} \bar h,
    \end{split}
\end{equation}
which provides the algebraic definition of the highest-weight states associated with the BMS$_4$ algebra. Unlike the case in BMS$_3$, where the primary states are still the eigenstates of $M_0$ with identical eigenvalues,  no primary state is the eigenstate of supertranslation generators $M_{r, s}$ in BMS$_4$. This is because acting any generator $M_{r, s}$ on a primary state $\ket{h, \bar h}$ shifts the conformal weights to $(h -r, \bar h -s)$, which is always different from $(h, \bar h)$ since $r, s$ can only take half-integer values. In other words, there is no generator $M_{r,s}$ commuting with $L_0, \bar{L}_0$ in BMS$_4$.

The descendants of the primary state $\ket{h, \bar h}$ can be expressed as
\begin{equation} \label{eq:des2D}
    \begin{split}
        & L_{-k_1} \cdots L_{-k_c} \bar L_{-j_1} \cdots \bar L_{-j_b} M_{- (n_1 -\frac12), - (\bar n_1 -\frac12)} \cdots M_{- (n_a -\frac12), - (\bar n_a -\frac12)} \ket{h, \bar h}; \\
        & (n_i \geq 1, ~ \text{and}~ \bar n_i \geq 1), \qquad (1 \leq j_1 \leq \cdots \leq j_b), \quad (1 \leq k_1 \leq \cdots \leq k_c).
    \end{split}
\end{equation}
Here, $a,b,c$ can be any non-negative integers, and $a = b = c = 0$ denotes the primary state $\ket{h, \bar h}$. All these states are the eigenstates of $L_0$ and $ \bar L_0$, with the eigenvalues $h + k_1 + \cdots + k_c + n_1 + \cdots + n_a - \frac{a}{2} \geq h$ and $\bar h + j_1 + \cdots + j_b + \bar n_1 + \cdots + \bar n_a - \frac{a}{2} \geq \bar h$, respectively. The equality holds if and only if the state is primary. Therefore, the primary state $\ket{h, \bar h}$ and its descendants span an irreducible HWR of BMS$_4$, or the highest-weight BMS$_4$ module $\mathcal{V} (h, \bar h)$.

In a more general construction, such as the 3D Carrollian CFT discussed in \cite{Chen:2021xkw}, the supertranslations exhibit a richer structure, which leads to the appearance of multiplet primary states in the theory. In such cases, supertranslations can act in multiple ways: they can change the conformal weight of a primary state, turning it into a descendant; they can also map between different components within a multiplet primary state. Therefore, the realization in the present work corresponds precisely to the singlet case. The key difference arises because on $S^2$ supertranslations act only as weight-shifting operators, while on higher-dimensional manifolds,  they can also generate spacetime coordinate transformations, similar to the discussion in the BMS$_3$ case.

The highest-weight vacuum of BMS$_4$ is not the primary state $\ket{h = 0, \bar h = 0}$, due to the difference in their annihilation conditions \eqref{eq:ML0=0} and \eqref{eq:LMh0}. The descendants of the vacuum state can be expressed as
\begin{equation}
    \begin{split}
        & L_{-k_1} \cdots L_{-k_c} \bar L_{-j_1} \cdots \bar L_{-j_b} M_{- (n_1 -\frac12), - (\bar n_1 -\frac12)} \cdots M_{- (n_a -\frac12), - (\bar n_a -\frac12)} \ket{\text{vac}}; \\
        & (n_i \geq 2 ~ \text{and} ~ \bar n_i \geq 2), \qquad (2 \leq j_1 \leq \cdots \leq j_b), \, (2 \leq k_1 \leq \cdots \leq k_c).
    \end{split}
\end{equation}
The vacuum state $\ket{\text{vac}}$ and its descendants span an irreducible representation of BMS$_4$, or the vacuum module $\mathcal{V}_{\text{vac}}$. Similarly, $\mathcal{V}_{\text{vac}}$ also serves as a HWR as $\ket{\text{vac}}$ has the smallest weights $h = \bar h =0$ compared to other states in the vacuum module.

The inner product of the states within the same module can be non-vanishing only when the two states are at the same level. However,  the states at the same level can still have vanishing inner products. For example, the descendants that are only generated from supertranslation are all zero-norm states, which can be seen directly from the annihilation conditions \eqref{eq:ML0=0}, \eqref{eq:ML0=0+} and \eqref{eq:LMh0}, \eqref{eq:LMh0+}. Furthermore, these zero-norm states do not satisfy the highest-weight conditions \eqref{eq:LMh0}, indicating that our constructions of BMS$_4$ HWRs on $S^2$ are generally non-unitary.

\subsection{Characters of the highest-weight modules}

The character of a highest-weight module $\mathcal{V} (h, \bar h)$ of BMS$_4$ on $S^2$ can be expressed as
\begin{equation}
    \chi_{h, \bar h} ({\bf{a}},{\bf{\bar a}},{\bf{b}}) = \text{tr}_{\mathcal{V}} \left[ \text{exp} \left( \sum_{i \in \mathbb Z} a_i L_i + \sum_{j \in \mathbb Z} \bar a_j \bar L_j + \sum_{m,n \in \mathbb Z} b_{m,n} M_{m - \frac{1}{2}, n - \frac{1}{2}} \right) \right],
\end{equation}
where ${\bf{a}}=\{a_0,\cdots,a_i,\cdots\},\,{\bf{\bar a}}=\{\bar a_0,\cdots,\bar a_j,\cdots\},{\bf{b}}=\{b_{0,0},\cdots,b_{m,n},\cdots\}$ are some complex variables. Note that only $L_0, \bar L_0$ contribute because the diagonal elements of other modes all vanish. Here, the supertranslation mode does not contribute, as there is no mode similar to $M_0$ in BMS$_3$ case. Conventionally, we choose $a_0 = \alpha, \, \bar a_0 = \bar \alpha$. Therefore, the character is
\begin{equation}
        \chi_{h, \bar h} (\alpha, \bar \alpha)  = \text{tr}_{\mathcal{V}} \left( e^{- \alpha L_0} e^{ - \bar \alpha \bar L_0} \right) = \sum_{h_{\mathcal{V}} \geq h, \, \bar h_{\mathcal{V}} \geq \bar h}  \text{dim}_{(h_{\mathcal{V}}, \bar h_{\mathcal{V}})} \, e^{- \alpha h_{\mathcal{V}} - \bar \alpha \bar h_{\mathcal{V}}},
\end{equation}
where dim$_{(h_{\mathcal{V}}, \bar h_{\mathcal{V}})}$ denotes the number of the weight-$(h_{\mathcal{V}}, \bar h_{\mathcal{V}})$ states within the module $\mathcal{V} (h, \bar h)$.

\begin{figure}[htbp]
    \centering
    \includegraphics[width=0.9\linewidth]{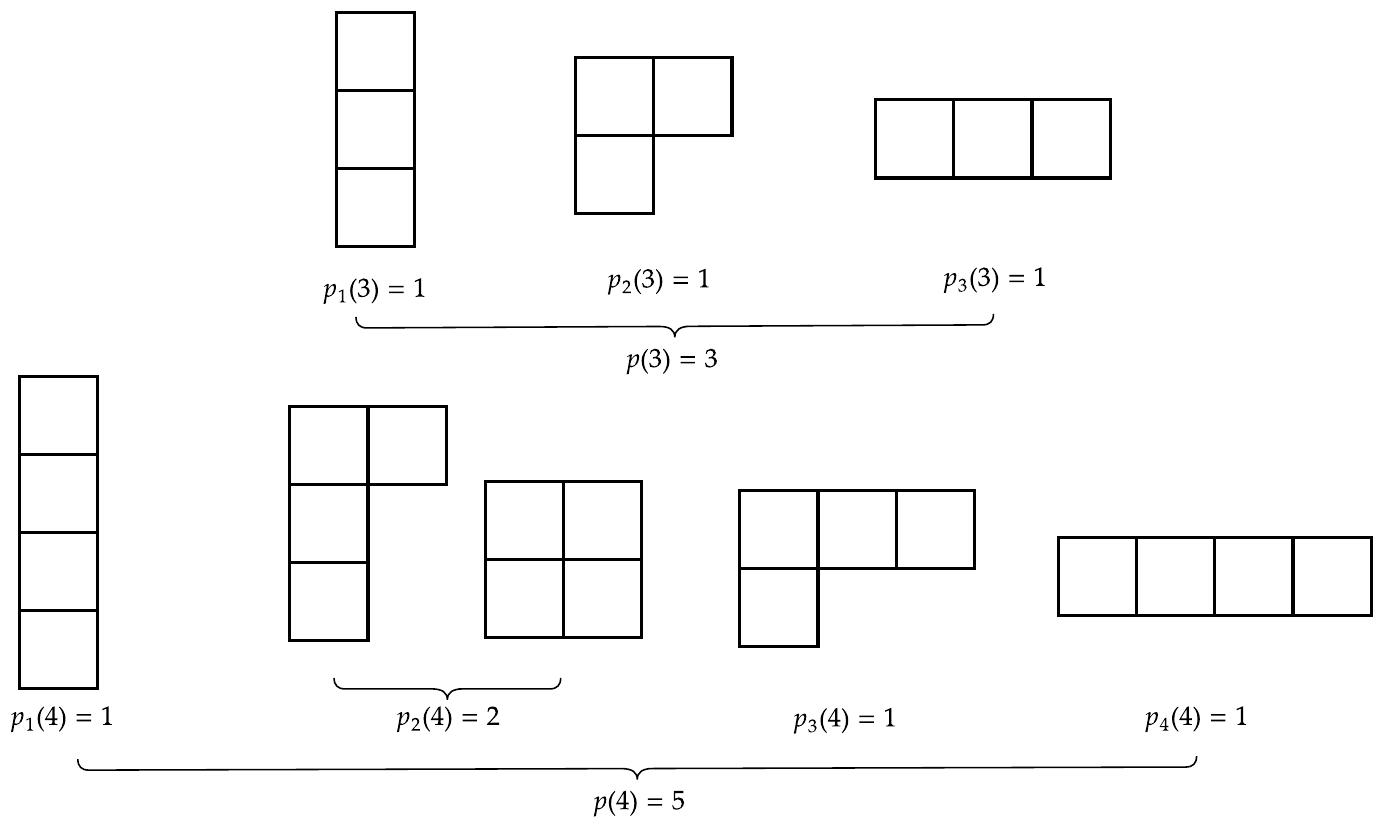}
    \caption{Counting $p_a (n)$ and $p(n)$ through Young diagram. Take $n = 3, 4$ as examples.}
    \label{fig:Young}
\end{figure}

The general descendants \eqref{eq:des2D} that share the same conformal weights $(h_{\mathcal{V}}, \bar h_{\mathcal{V}}) = (h + n_L + n_M - \frac{a}{2}, \bar h + \bar n_{\bar L} + \bar n_M - \frac{a}{2})$ have $p_c (n_L) p_b (\bar n_{\bar L}) p_a (n_M) p_a (\bar n_M)$ choices. Here, $n_L = k_1 + \cdots + k_c, \, \bar n_{\bar L} = j_1 + \cdots + j_b, \, n_M = n_1 + \cdots + n_a, \, \bar n_M = \bar n_1 + \cdots + \bar n_a$ and $a,b,c$ are fixed but $k_i, j_i, m_i, n_i$ are variables. $p_k (n)$ denotes the number of ways to partition the positive integer $n$ into $k$ parts of positive integers, or more explicitly, $p_k (n)$ counts the number of Young diagrams with length (number of parts) $k$ and size (total number of boxes) $n$. Furthermore, the number of size-$n$ Young diagrams can be counted by
\begin{equation}
    \sum_{a = 1}^n p_a (n) = p (n),
\end{equation}
where $p (n)$ is given by \eqref{eq:Euler}. To be more explicit, we illustrate the process of calculating $p_a (n)$ and $p (n)$ using Young diagrams in Fig.\ref{fig:Young}, with examples $n = 3, 4$. The partition $p_a (n)$ can be generated through the following function
\begin{equation}
    \sum_{n = k}^\infty p_k (n) x^n = \frac{x^k}{\prod_{i = 1}^k (1-x^i)} \qquad (\forall k \in \mathbb Z_{>0}).
\end{equation}
For more information on the partitions $p(n)$ and $p_a(n)$, we refer to \cite{andrews1998theory}.

The character can then be derived as
\begin{multline} \label{eq:chi2Dbeg}
    \chi_{h, \bar h} (\alpha, \bar \alpha) = \sum_{n_L, \bar n_L, n_M, \bar n_M=0}^\infty \sum_{c=0}^{n_L} \sum_{b=0}^{\bar n_{\bar L}} \sum_{a=0}^{\text{min} \{n_M, \bar n_M\}} p_c (n_L) p_b (\bar n_{\bar L}) p_a (n_M) p_a (\bar n_M) \\
    \times e^{-\alpha (h + n_L + n_M - \frac{a}{2})} e^{- \bar \alpha (\bar h + \bar n_{\bar L} + \bar n_M - \frac{a}{2})}.
\end{multline}
It is important to note that $p_0 (n)=0$ for $n>0$, which can also be seen in Fig.\ref{fig:Young}. Moreover, we need to introduce the conventional choice $p_0 (0)=1$. A quick remark is that the supertranslation modes vary both $h$ and $\bar h$. The effects on both parts cannot be factorized. Hence, the upper bound in the summation of index $a$ should be the smaller one of $n_M$ and $\bar n_M$ since the number of the supertranslation modes should not be larger than either of the conformal weights that are shifted by the supertranslations. Finally, the character of a highest-weight module $\mathcal{V} (h, \bar h)$ is factorized into the Virasoro parts and the supertranslation part
\begin{multline} \label{eq:chi2dDs}
    \chi_{h, \bar h} (\alpha, \bar \alpha) 
    = \frac{e^{- \alpha h - \bar \alpha \bar h}}{\prod^\infty_{k = 1} (1- e^{- \alpha k}) (1- e^{- \bar \alpha k})} \varphi^M (e^{-\alpha}, e^{-\bar \alpha}) \\
    = \frac{e^{- \alpha h - \bar \alpha \bar h}}{\prod^\infty_{k = 1} (1- e^{- \alpha k}) (1- e^{- \bar \alpha k})} \left[ 1+ \sum_{a=1}^\infty \frac{e^{- a \frac{\alpha + \bar \alpha}{2}}}{\prod^a_{j=1} (1- e^{- \alpha j}) (1- e^{- \bar \alpha j})} \right],
\end{multline}
where the contribution of supertranslation is expressed by the function $\varphi^M (e^{-\alpha}, e^{-\bar \alpha})$, which can be computed as
\begin{equation}
    \begin{split}
        \varphi^M (x, y) & = \sum_{n_M, \bar n_M = 0}^\infty \sum_{a = 0}^{\text{min} \{n_M, \bar n_M\}} p_a (n_M) p_a (\bar n_M) x^{n_M - \frac{a}{2}} y^{\bar n_M - \frac{a}{2}} \\
        & = \sum_{a=0}^\infty (x y)^{-\frac{a}{2}} \sum_{n_M = a}^\infty p_a (n_M)  x^{n_M} \sum_{\bar n_M = a}^\infty p_a (\bar n_M) y^{\bar n_M} \\
        & = 1 + \sum_{a=1}^\infty \frac{(xy)^{\frac{a}{2}}}{\prod^a_{i=1} (1-x^i) (1-y^i)}.
    \end{split}
\end{equation}
The factorization of supertranslation and superrotation descendants in the character is reasonable, as supertranslation does not introduce a new quantum number that changes the highest-weight state.

Similar to the discussion in BMS$_3$, the character of the vacuum module can be derived by subtracting the contributions of the descendants contain $L_{-1}, \bar L_{-1}, M_{- (n - \frac12), - (\bar n - \frac12)}$ $(n=1 ~\text{or} ~ \bar n = 1)$ of the $h = 0, \bar h = 0$ case in \eqref{eq:chi2dDs}, which produces the factor $(1 - e^{- \alpha})^2 (1 - e^{- \bar \alpha})^2$. 
Consequently, one obtains the character of the vacuum module
\begin{equation} \label{eq:vacchi2d}
    \begin{split}
        \chi_{\text{vac}} (\alpha, \bar \alpha) = & (1 - e^{- \alpha})^2 (1 - e^{- \bar \alpha})^2 \chi_{0, 0} (\alpha, \bar \alpha) \\
        = & \left[ (1 - e^{- \alpha}) (1 - e^{- \bar \alpha}) + e^{- \frac{\alpha + \bar \alpha}{2}} + \sum_{a=2}^\infty \frac{e^{- a \frac{\alpha + \bar \alpha}{2}}}{\prod^a_{j=2} (1- e^{- \alpha j}) (1- e^{- \bar \alpha j})}\right] \\
        & \times \frac{1}{\prod^\infty_{k = 2} (1- e^{- \alpha k}) (1- e^{- \bar \alpha k})}.
    \end{split}
\end{equation}
The contributions of supertranslation are in the brackets in the second line, which are decomposed from the Virasoro contributions in the last line. Note that the Virasoro contribution is the same as the character of the AdS$_3$ gravity \cite{Witten:2007kt,Maloney:2007ud,Giombi:2008vd,David:2009xg} with vanishing central charge; the main difference lies in the existence of supertranslations.

Given that four-dimensional gravity possesses propagating degrees of freedom, the highest-weight states besides the vacuum of the theory on the codimension-two boundary are expected to have the gravitational dual in the bulk of Einstein theory. It is reasonable to expect that the 4D gravitational partition function is related to the characters \eqref{eq:vacchi2d} and \eqref{eq:chi2dDs}.

\section{Conclusions and Discussions} \label{sec:conclusion}

In this work, we investigated the highest weight representation of the BMS$_{d+2}$ algebra within the framework of the BMS$_{d+2}$-invariant field theory on the codimension-two boundary $S^d$ of AFS$_{d+2}$ for $d = 1, 2$. The construction presents some distinct features. Most importantly, the supertranslations now act as the weight shifting rather than more general coordinate transformations in the Carrollian case. As a result, the primary states are now characterized solely by their conformal weights, with either a fixed charge or no charge. This is the most remarkable feature in the present construction.

Furthermore, we computed the BMS characters for the highest-weight modules, including the HWR constructed from the primary states and their descendants, as well as the vacuum state and its descendants. In particular, the BMS$_3$ character of the vacuum module on $S^1$, for an appropriate choice of central charges, coincides with the one-loop partition function of three-dimensional Einstein gravity expanded around the Minkowski background, up to an overall imaginary factor ``$i$''. This provides valuable insights into the 4D gravitational partition function. We expect that the BMS$_4$ highest-weight characters could be dual to the partition functions for the Einstein gravity fluctuated near a 4D standard Minkowski vacuum \cite{chen2025}.

\acknowledgments

The authors thank Glenn Barnich, Zezhou Hu, Yi Li, Hao Ouyang, Yu-ting Wen, Jie Xu, Zhi-jun Yin, and Yu-fan Zheng for the valuable discussions. This research was supported in part by the National Natural Science Foundation of China (NSFC) under Grants No.~11735001, No.~12275004, No.~12475053, No.~12235016, No.~11935009. P.M.~is also supported in part by Tianjin University Self-Innovation Fund Extreme Basic Research Project Grant No. 2025XJ21-0007.

\bibliographystyle{JHEP}
\bibliography{biblio.bib}

\providecommand{\href}[2]{#2}\begingroup\raggedright\begin{thebibliography}{10}

\bibitem{Noether:1918zz}
E.~Noether, \emph{{Invariant Variation Problems}},
  \href{https://doi.org/10.1080/00411457108231446}{\emph{Gott. Nachr.}
  {\bfseries 1918} (1918) 235}
  [\href{https://arxiv.org/abs/physics/0503066}{{\ttfamily physics/0503066}}].

\bibitem{Witten:2007kt}
E.~Witten, \emph{{Three-Dimensional Gravity Revisited}},
  \href{https://arxiv.org/abs/0706.3359}{{\ttfamily 0706.3359}}.

\bibitem{Maloney:2007ud}
A.~Maloney and E.~Witten, \emph{{Quantum Gravity Partition Functions in Three
  Dimensions}}, \href{https://doi.org/10.1007/JHEP02(2010)029}{\emph{JHEP}
  {\bfseries 02} (2010) 029} [\href{https://arxiv.org/abs/0712.0155}{{\ttfamily
  0712.0155}}].

\bibitem{Giombi:2008vd}
S.~Giombi, A.~Maloney and X.~Yin, \emph{{One-loop Partition Functions of 3D
  Gravity}}, \href{https://doi.org/10.1088/1126-6708/2008/08/007}{\emph{JHEP}
  {\bfseries 08} (2008) 007} [\href{https://arxiv.org/abs/0804.1773}{{\ttfamily
  0804.1773}}].

\bibitem{David:2009xg}
J.R.~David, M.R.~Gaberdiel and R.~Gopakumar, \emph{{The Heat Kernel on AdS(3)
  and its Applications}},
  \href{https://doi.org/10.1007/JHEP04(2010)125}{\emph{JHEP} {\bfseries 04}
  (2010) 125} [\href{https://arxiv.org/abs/0911.5085}{{\ttfamily 0911.5085}}].

\bibitem{Barnich:2015mui}
G.~Barnich, H.A.~Gonzalez, A.~Maloney and B.~Oblak, \emph{{One-loop partition
  function of three-dimensional flat gravity}},
  \href{https://doi.org/10.1007/JHEP04(2015)178}{\emph{JHEP} {\bfseries 04}
  (2015) 178} [\href{https://arxiv.org/abs/1502.06185}{{\ttfamily
  1502.06185}}].

\bibitem{Oblak:2015sea}
B.~Oblak, \emph{{Characters of the BMS Group in Three Dimensions}},
  \href{https://doi.org/10.1007/s00220-015-2408-7}{\emph{Commun. Math. Phys.}
  {\bfseries 340} (2015) 413}
  [\href{https://arxiv.org/abs/1502.03108}{{\ttfamily 1502.03108}}].

\bibitem{Brown:1986nw}
J.D.~Brown and M.~Henneaux, \emph{{Central Charges in the Canonical Realization
  of Asymptotic Symmetries: An Example from Three-Dimensional Gravity}},
  \href{https://doi.org/10.1007/BF01211590}{\emph{Commun. Math. Phys.}
  {\bfseries 104} (1986) 207}.

\bibitem{Ashtekar:1996cd}
A.~Ashtekar, J.~Bicak and B.G.~Schmidt, \emph{{Asymptotic structure of symmetry
  reduced general relativity}},
  \href{https://doi.org/10.1103/PhysRevD.55.669}{\emph{Phys. Rev. D} {\bfseries
  55} (1997) 669} [\href{https://arxiv.org/abs/gr-qc/9608042}{{\ttfamily
  gr-qc/9608042}}].

\bibitem{Barnich:2006av}
G.~Barnich and G.~Compere, \emph{{Classical central extension for asymptotic
  symmetries at null infinity in three spacetime dimensions}},
  \href{https://doi.org/10.1088/0264-9381/24/5/F01}{\emph{Class. Quant. Grav.}
  {\bfseries 24} (2007) F15}
  [\href{https://arxiv.org/abs/gr-qc/0610130}{{\ttfamily gr-qc/0610130}}].

\bibitem{Bondi:1962px}
H.~Bondi, M.G.J.~van~der Burg and A.W.K.~Metzner, \emph{{Gravitational waves in
  general relativity. 7. Waves from axisymmetric isolated systems}},
  \href{https://doi.org/10.1098/rspa.1962.0161}{\emph{Proc. Roy. Soc. Lond. A}
  {\bfseries 269} (1962) 21}.

\bibitem{Sachs:1962wk}
R.K.~Sachs, \emph{{Gravitational waves in general relativity. 8. Waves in
  asymptotically flat space-times}},
  \href{https://doi.org/10.1098/rspa.1962.0206}{\emph{Proc. Roy. Soc. Lond. A}
  {\bfseries 270} (1962) 103}.

\bibitem{Sachs:1962zza}
R.~Sachs, \emph{{Asymptotic symmetries in gravitational theory}},
  \href{https://doi.org/10.1103/PhysRev.128.2851}{\emph{Phys. Rev.} {\bfseries
  128} (1962) 2851}.

\bibitem{Barnich:2009se}
G.~Barnich and C.~Troessaert, \emph{{Symmetries of asymptotically flat 4
  dimensional spacetimes at null infinity revisited}},
  \href{https://doi.org/10.1103/PhysRevLett.105.111103}{\emph{Phys. Rev. Lett.}
  {\bfseries 105} (2010) 111103}
  [\href{https://arxiv.org/abs/0909.2617}{{\ttfamily 0909.2617}}].

\bibitem{Barnich:2010ojg}
G.~Barnich and C.~Troessaert, \emph{{Supertranslations call for
  superrotations}}, \href{https://doi.org/10.22323/1.127.0010}{\emph{PoS}
  {\bfseries CNCFG2010} (2010) 010}
  [\href{https://arxiv.org/abs/1102.4632}{{\ttfamily 1102.4632}}].

\bibitem{Barnich:2010eb}
G.~Barnich and C.~Troessaert, \emph{{Aspects of the BMS/CFT correspondence}},
  \href{https://doi.org/10.1007/JHEP05(2010)062}{\emph{JHEP} {\bfseries 05}
  (2010) 062} [\href{https://arxiv.org/abs/1001.1541}{{\ttfamily 1001.1541}}].

\bibitem{Strominger:2017zoo}
A.~Strominger, \emph{{Lectures on the Infrared Structure of Gravity and Gauge
  Theory}}, Princeton University Press, Princeton (2018),
  [\href{https://arxiv.org/abs/1703.05448}{{\ttfamily 1703.05448}}].

\bibitem{Pasterski:2016qvg}
S.~Pasterski, S.-H.~Shao and A.~Strominger, \emph{{Flat Space Amplitudes and
  Conformal Symmetry of the Celestial Sphere}},
  \href{https://doi.org/10.1103/PhysRevD.96.065026}{\emph{Phys. Rev. D}
  {\bfseries 96} (2017) 065026}
  [\href{https://arxiv.org/abs/1701.00049}{{\ttfamily 1701.00049}}].

\bibitem{Pasterski:2017kqt}
S.~Pasterski and S.-H.~Shao, \emph{{Conformal basis for flat space
  amplitudes}}, \href{https://doi.org/10.1103/PhysRevD.96.065022}{\emph{Phys.
  Rev. D} {\bfseries 96} (2017) 065022}
  [\href{https://arxiv.org/abs/1705.01027}{{\ttfamily 1705.01027}}].

\bibitem{Pasterski:2017ylz}
S.~Pasterski, S.-H.~Shao and A.~Strominger, \emph{{Gluon Amplitudes as 2d
  Conformal Correlators}},
  \href{https://doi.org/10.1103/PhysRevD.96.085006}{\emph{Phys. Rev. D}
  {\bfseries 96} (2017) 085006}
  [\href{https://arxiv.org/abs/1706.03917}{{\ttfamily 1706.03917}}].

\bibitem{Raclariu:2021zjz}
A.-M.~Raclariu, \emph{{Lectures on Celestial Holography}},
  \href{https://arxiv.org/abs/2107.02075}{{\ttfamily 2107.02075}}.

\bibitem{Pasterski:2021rjz}
S.~Pasterski, \emph{{Lectures on celestial amplitudes}},
  \href{https://doi.org/10.1140/epjc/s10052-021-09846-7}{\emph{Eur. Phys. J. C}
  {\bfseries 81} (2021) 1062}
  [\href{https://arxiv.org/abs/2108.04801}{{\ttfamily 2108.04801}}].

\bibitem{Pasterski:2021raf}
S.~Pasterski, M.~Pate and A.-M.~Raclariu, \emph{{Celestial Holography}},  in
  \emph{{Snowmass 2021}}, 11, 2021
  [\href{https://arxiv.org/abs/2111.11392}{{\ttfamily 2111.11392}}].

\bibitem{Donnay:2022aba}
L.~Donnay, A.~Fiorucci, Y.~Herfray and R.~Ruzziconi, \emph{{Carrollian
  Perspective on Celestial Holography}},
  \href{https://doi.org/10.1103/PhysRevLett.129.071602}{\emph{Phys. Rev. Lett.}
  {\bfseries 129} (2022) 071602}
  [\href{https://arxiv.org/abs/2202.04702}{{\ttfamily 2202.04702}}].

\bibitem{Donnay:2022wvx}
L.~Donnay, A.~Fiorucci, Y.~Herfray and R.~Ruzziconi, \emph{{Bridging Carrollian
  and celestial holography}},
  \href{https://doi.org/10.1103/PhysRevD.107.126027}{\emph{Phys. Rev. D}
  {\bfseries 107} (2023) 126027}
  [\href{https://arxiv.org/abs/2212.12553}{{\ttfamily 2212.12553}}].

\bibitem{Bagchi:2022emh}
A.~Bagchi, S.~Banerjee, R.~Basu and S.~Dutta, \emph{{Scattering Amplitudes:
  Celestial and Carrollian}},
  \href{https://doi.org/10.1103/PhysRevLett.128.241601}{\emph{Phys. Rev. Lett.}
  {\bfseries 128} (2022) 241601}
  [\href{https://arxiv.org/abs/2202.08438}{{\ttfamily 2202.08438}}].

\bibitem{Chen:2023naw}
B.~Chen and Z.~Hu, \emph{{Bulk reconstruction in flat holography}},
  \href{https://doi.org/10.1007/JHEP03(2024)064}{\emph{JHEP} {\bfseries 03}
  (2024) 064} [\href{https://arxiv.org/abs/2312.13574}{{\ttfamily
  2312.13574}}].

\bibitem{Chen:2023pqf}
B.~Chen, R.~Liu, H.~Sun and Y.-f.~Zheng, \emph{{Constructing Carrollian field
  theories from null reduction}},
  \href{https://doi.org/10.1007/JHEP11(2023)170}{\emph{JHEP} {\bfseries 11}
  (2023) 170} [\href{https://arxiv.org/abs/2301.06011}{{\ttfamily
  2301.06011}}].

\bibitem{Bagchi:2023cen}
A.~Bagchi, P.~Dhivakar and S.~Dutta, \emph{{Holography in flat spacetimes: the
  case for Carroll}},
  \href{https://doi.org/10.1007/JHEP08(2024)144}{\emph{JHEP} {\bfseries 08}
  (2024) 144} [\href{https://arxiv.org/abs/2311.11246}{{\ttfamily
  2311.11246}}].

\bibitem{RLINA_1967_8_43_1-2_a5}
V.~Cantoni, \emph{Induction of representations of the generalized
  {Bondi-Metzner} group}, {\emph{Atti della Accademia nazionale dei Lincei.
  Rendiconti della Classe di scienze fisiche, matematiche e naturali}
  {\bfseries Ser. 8, 43} (1967) 30}.

\bibitem{cantoni1966class}
V.~Cantoni, \emph{A class of representations of the generalized bondi—metzner
  group}, {\emph{Journal of Mathematical Physics} {\bfseries 7} (1966) 1361}.

\bibitem{cantoni1967reduction}
V.~Cantoni, \emph{Reduction of some representations of the generalized
  bondi-metzner group}, {\emph{Journal of Mathematical Physics} {\bfseries 8}
  (1967) 1700}.

\bibitem{Crampin:1974aw}
M.~Crampin, \emph{{Physical significance of the topology of the
  bondi-metzner-sachs group}},
  \href{https://doi.org/10.1103/PhysRevLett.33.547}{\emph{Phys. Rev. Lett.}
  {\bfseries 33} (1974) 547}.

\bibitem{Girardello:1974sq}
L.~Girardello and G.~Parravicini, \emph{{Continuous spins in the
  bondi-metzner-sachs group of asymptotic symmetry in general relativity}},
  \href{https://doi.org/10.1103/PhysRevLett.32.565}{\emph{Phys. Rev. Lett.}
  {\bfseries 32} (1974) 565}.

\bibitem{mccarthy1}
P.J.~McCarthy, \emph{Representations of the bondi—metzner—sachs group i.
  determination of the representations}, {\emph{Proceedings of the Royal
  Society of London. A. Mathematical and Physical Sciences} {\bfseries 330}
  (1972) 517}.

\bibitem{mccarthy2}
P.J.~McCarthy, \emph{Representations of the bondi-metzner-sachs group-ii.
  properties and classification of the representations}, {\emph{Proceedings of
  the Royal Society of London. A. Mathematical and Physical Sciences}
  {\bfseries 333} (1973) 317}.

\bibitem{mccarthy3}
P.J.~McCarthy and M.~Crampin, \emph{Representations of the
  bondi-metzner—sachs group iii. poincar{\'e} spin multiplicities and
  irreducibility}, {\emph{Proceedings of the Royal Society of London. A.
  Mathematical and Physical Sciences} {\bfseries 335} (1973) 301}.

\bibitem{mccarthy4}
M.~Crampin and P.J.~McCarthy, \emph{Representations of the bondi-metzner-sachs
  group iv. cantoni representations are induced}, {\emph{Proceedings of the
  Royal Society of London. A. Mathematical and Physical Sciences} {\bfseries
  351} (1976) 55}.

\bibitem{Mccarthy:1972ry}
P.J.M.~Mccarthy, \emph{{Asymptotically flat space-times and elementary
  particles}}, \href{https://doi.org/10.1103/PhysRevLett.29.817}{\emph{Phys.
  Rev. Lett.} {\bfseries 29} (1972) 817}.

\bibitem{mccarthy1978lifting}
P.J.~McCarthy, \emph{Lifting of projective representations of the
  bondi—metzner—sachs group}, {\emph{Proceedings of the Royal Society of
  London. A. Mathematical and Physical Sciences} {\bfseries 358} (1978) 141}.

\bibitem{mccarthy1975bondi}
P.J.~McCarthy, \emph{The bondi—metzner—sachs group in the nuclear
  topology}, {\emph{Proceedings of the Royal Society of London. A. Mathematical
  and Physical Sciences} {\bfseries 343} (1975) 489}.

\bibitem{mccarthy78hyp}
P.J.~McCarthy, \emph{Hyperfunctions and asymptotic symmetries},
  {\emph{Proceedings of the Royal Society of London. A. Mathematical and
  Physical Sciences} {\bfseries 358} (1978) 495}.

\bibitem{Bekaert:2024uuy}
X.~Bekaert, L.~Donnay and Y.~Herfray, \emph{{BMS particles}},
  \href{https://doi.org/10.1103/8376-fync}{\emph{Phys. Rev. Lett.} {\bfseries
  135} (2025) 131602} [\href{https://arxiv.org/abs/2412.06002}{{\ttfamily
  2412.06002}}].

\bibitem{Bekaert:2025kjb}
X.~Bekaert and Y.~Herfray, \emph{{BMS representations for generic
  supermomentum}},  \href{https://arxiv.org/abs/2505.05368}{{\ttfamily
  2505.05368}}.

\bibitem{Prabhu:2022zcr}
K.~Prabhu, G.~Satishchandran and R.M.~Wald, \emph{{Infrared finite scattering
  theory in quantum field theory and quantum gravity}},
  \href{https://doi.org/10.1103/PhysRevD.106.066005}{\emph{Phys. Rev. D}
  {\bfseries 106} (2022) 066005}
  [\href{https://arxiv.org/abs/2203.14334}{{\ttfamily 2203.14334}}].

\bibitem{Prabhu:2024lmg}
K.~Prabhu and G.~Satishchandran, \emph{{Infrared finite scattering theory:
  Amplitudes and soft theorems}},
  \href{https://doi.org/10.1103/PhysRevD.110.085022}{\emph{Phys. Rev. D}
  {\bfseries 110} (2024) 085022}
  [\href{https://arxiv.org/abs/2402.18637}{{\ttfamily 2402.18637}}].

\bibitem{Prabhu:2024zwl}
K.~Prabhu and G.~Satishchandran, \emph{{Infrared finite scattering theory:
  scattering states and representations of the BMS group}},
  \href{https://doi.org/10.1007/JHEP08(2024)055}{\emph{JHEP} {\bfseries 08}
  (2024) 055} [\href{https://arxiv.org/abs/2402.00102}{{\ttfamily
  2402.00102}}].

\bibitem{Barnich:2021dta}
G.~Barnich and R.~Ruzziconi, \emph{{Coadjoint representation of the BMS group
  on celestial Riemann surfaces}},
  \href{https://doi.org/10.1007/JHEP06(2021)079}{\emph{JHEP} {\bfseries 06}
  (2021) 079} [\href{https://arxiv.org/abs/2103.11253}{{\ttfamily
  2103.11253}}].

\bibitem{Barnich:2022bni}
G.~Barnich, K.~Nguyen and R.~Ruzziconi, \emph{{Geometric action for extended
  Bondi-Metzner-Sachs group in four dimensions}},
  \href{https://doi.org/10.1007/JHEP12(2022)154}{\emph{JHEP} {\bfseries 12}
  (2022) 154} [\href{https://arxiv.org/abs/2211.07592}{{\ttfamily
  2211.07592}}].

\bibitem{AliAhmad:2025hdl}
S.~Ali~Ahmad, \emph{{The building blocks of asymptotically flat spacetimes}},
  \href{https://arxiv.org/abs/2506.11190}{{\ttfamily 2506.11190}}.

\bibitem{Barnich:2014kra}
G.~Barnich and B.~Oblak, \emph{{Notes on the BMS group in three dimensions: I.
  Induced representations}},
  \href{https://doi.org/10.1007/JHEP06(2014)129}{\emph{JHEP} {\bfseries 06}
  (2014) 129} [\href{https://arxiv.org/abs/1403.5803}{{\ttfamily 1403.5803}}].

\bibitem{Barnich:2015uva}
G.~Barnich and B.~Oblak, \emph{{Notes on the BMS group in three dimensions: II.
  Coadjoint representation}},
  \href{https://doi.org/10.1007/JHEP03(2015)033}{\emph{JHEP} {\bfseries 03}
  (2015) 033} [\href{https://arxiv.org/abs/1502.00010}{{\ttfamily
  1502.00010}}].

\bibitem{Oblak:2016eij}
B.~Oblak, \emph{{BMS Particles in Three Dimensions}}, Ph.D. thesis, U.
  Brussels, Brussels U., 2016.
\newblock \href{https://arxiv.org/abs/1610.08526}{{\ttfamily 1610.08526}}.
\newblock 10.1007/978-3-319-61878-4.

\bibitem{Campoleoni:2016vsh}
A.~Campoleoni, H.A.~Gonzalez, B.~Oblak and M.~Riegler, \emph{{BMS Modules in
  Three Dimensions}},
  \href{https://doi.org/10.1142/S0217751X16500688}{\emph{Int. J. Mod. Phys. A}
  {\bfseries 31} (2016) 1650068}
  [\href{https://arxiv.org/abs/1603.03812}{{\ttfamily 1603.03812}}].

\bibitem{Melas:2017ggm}
E.~Melas, \emph{{Representations of the
  Bondi\textemdash{}Metzner\textemdash{}Sachs group in three
  space\textemdash{}time dimensions}},
  \href{https://doi.org/10.1088/1742-6596/804/1/012030}{\emph{J. Phys. Conf.
  Ser.} {\bfseries 804} (2017) 012030}.

\bibitem{Melas:2017jzb}
E.~Melas, \emph{{On the representation theory of the
  Bondi\textendash{}Metzner\textendash{}Sachs group and its variants in three
  space\textendash{}time dimensions}},
  \href{https://doi.org/10.1063/1.4993198}{\emph{J. Math. Phys.} {\bfseries 58}
  (2017) 071705} [\href{https://arxiv.org/abs/1703.05980}{{\ttfamily
  1703.05980}}].

\bibitem{Melas:2021oje}
E.~Melas, \emph{{Representations of the Bondi-Metzner-Sachs group in three
  space-time dimensions in the Hilbert topology I. Determination of the
  representations}},  \href{https://arxiv.org/abs/2108.00424}{{\ttfamily
  2108.00424}}.

\bibitem{Stieberger:2018onx}
S.~Stieberger and T.R.~Taylor, \emph{{Symmetries of Celestial Amplitudes}},
  \href{https://doi.org/10.1016/j.physletb.2019.03.063}{\emph{Phys. Lett. B}
  {\bfseries 793} (2019) 141}
  [\href{https://arxiv.org/abs/1812.01080}{{\ttfamily 1812.01080}}].

\bibitem{Fotopoulos:2019vac}
A.~Fotopoulos, S.~Stieberger, T.R.~Taylor and B.~Zhu, \emph{{Extended BMS
  Algebra of Celestial CFT}},
  \href{https://doi.org/10.1007/JHEP03(2020)130}{\emph{JHEP} {\bfseries 03}
  (2020) 130} [\href{https://arxiv.org/abs/1912.10973}{{\ttfamily
  1912.10973}}].

\bibitem{Hao:2021urq}
P.-x.~Hao, W.~Song, X.~Xie and Y.~Zhong, \emph{{BMS-invariant free scalar
  model}}, \href{https://doi.org/10.1103/PhysRevD.105.125005}{\emph{Phys. Rev.
  D} {\bfseries 105} (2022) 125005}
  [\href{https://arxiv.org/abs/2111.04701}{{\ttfamily 2111.04701}}].

\bibitem{Hao:2022xhq}
P.-X.~Hao, W.~Song, Z.~Xiao and X.~Xie, \emph{{BMS-invariant free fermion
  models}}, \href{https://doi.org/10.1103/PhysRevD.109.025002}{\emph{Phys. Rev.
  D} {\bfseries 109} (2024) 025002}
  [\href{https://arxiv.org/abs/2211.06927}{{\ttfamily 2211.06927}}].

\bibitem{Saha:2022gjw}
A.~Saha, \emph{{Intrinsic approach to 1 + 1D Carrollian Conformal Field
  Theory}}, \href{https://doi.org/10.1007/JHEP12(2022)133}{\emph{JHEP}
  {\bfseries 12} (2022) 133}
  [\href{https://arxiv.org/abs/2207.11684}{{\ttfamily 2207.11684}}].

\bibitem{Chen:2020vvn}
B.~Chen, P.-X.~Hao, R.~Liu and Z.-F.~Yu, \emph{{On Galilean conformal
  bootstrap}}, \href{https://doi.org/10.1007/JHEP06(2021)112}{\emph{JHEP}
  {\bfseries 06} (2021) 112}
  [\href{https://arxiv.org/abs/2011.11092}{{\ttfamily 2011.11092}}].

\bibitem{Chen:2022jhx}
B.~Chen, P.-x.~Hao, R.~Liu and Z.-f.~Yu, \emph{{On Galilean conformal
  bootstrap. Part II. \ensuremath{\xi} = 0 sector}},
  \href{https://doi.org/10.1007/JHEP12(2022)019}{\emph{JHEP} {\bfseries 12}
  (2022) 019} [\href{https://arxiv.org/abs/2207.01474}{{\ttfamily
  2207.01474}}].

\bibitem{Chen:2024voz}
B.~Chen, H.~Sun and Y.-f.~Zheng, \emph{{Quantization of Carrollian conformal
  scalar theories}},
  \href{https://doi.org/10.1103/PhysRevD.110.125010}{\emph{Phys. Rev. D}
  {\bfseries 110} (2024) 125010}
  [\href{https://arxiv.org/abs/2406.17451}{{\ttfamily 2406.17451}}].

\bibitem{Bagchi:2009pe}
A.~Bagchi, R.~Gopakumar, I.~Mandal and A.~Miwa, \emph{{GCA in 2d}},
  \href{https://doi.org/10.1007/JHEP08(2010)004}{\emph{JHEP} {\bfseries 08}
  (2010) 004} [\href{https://arxiv.org/abs/0912.1090}{{\ttfamily 0912.1090}}].

\bibitem{Bagchi:2009ca}
A.~Bagchi and I.~Mandal, \emph{{On Representations and Correlation Functions of
  Galilean Conformal Algebras}},
  \href{https://doi.org/10.1016/j.physletb.2009.04.030}{\emph{Phys. Lett. B}
  {\bfseries 675} (2009) 393}
  [\href{https://arxiv.org/abs/0903.4524}{{\ttfamily 0903.4524}}].

\bibitem{Bagchi:2016geg}
A.~Bagchi, M.~Gary and Zodinmawia, \emph{{Bondi-Metzner-Sachs bootstrap}},
  \href{https://doi.org/10.1103/PhysRevD.96.025007}{\emph{Phys. Rev. D}
  {\bfseries 96} (2017) 025007}
  [\href{https://arxiv.org/abs/1612.01730}{{\ttfamily 1612.01730}}].

\bibitem{Bagchi:2019unf}
A.~Bagchi, A.~Saha and Zodinmawia, \emph{{BMS Characters and Modular
  Invariance}}, \href{https://doi.org/10.1007/JHEP07(2019)138}{\emph{JHEP}
  {\bfseries 07} (2019) 138}
  [\href{https://arxiv.org/abs/1902.07066}{{\ttfamily 1902.07066}}].

\bibitem{Chen:2021xkw}
B.~Chen, R.~Liu and Y.-f.~Zheng, \emph{{On higher-dimensional Carrollian and
  Galilean conformal field theories}},
  \href{https://doi.org/10.21468/SciPostPhys.14.5.088}{\emph{SciPost Phys.}
  {\bfseries 14} (2023) 088}
  [\href{https://arxiv.org/abs/2112.10514}{{\ttfamily 2112.10514}}].

\bibitem{Barnich:2012xq}
G.~Barnich, \emph{{Entropy of three-dimensional asymptotically flat
  cosmological solutions}},
  \href{https://doi.org/10.1007/JHEP10(2012)095}{\emph{JHEP} {\bfseries 10}
  (2012) 095} [\href{https://arxiv.org/abs/1208.4371}{{\ttfamily 1208.4371}}].

\bibitem{Barnich:2012aw}
G.~Barnich, A.~Gomberoff and H.A.~Gonzalez, \emph{{The Flat limit of three
  dimensional asymptotically anti-de Sitter spacetimes}},
  \href{https://doi.org/10.1103/PhysRevD.86.024020}{\emph{Phys. Rev. D}
  {\bfseries 86} (2012) 024020}
  [\href{https://arxiv.org/abs/1204.3288}{{\ttfamily 1204.3288}}].

\bibitem{Barnich:2011mi}
G.~Barnich and C.~Troessaert, \emph{{BMS charge algebra}},
  \href{https://doi.org/10.1007/JHEP12(2011)105}{\emph{JHEP} {\bfseries 12}
  (2011) 105} [\href{https://arxiv.org/abs/1106.0213}{{\ttfamily 1106.0213}}].

\bibitem{Barnich:2017ubf}
G.~Barnich, \emph{{Centrally extended BMS4 Lie algebroid}},
  \href{https://doi.org/10.1007/JHEP06(2017)007}{\emph{JHEP} {\bfseries 06}
  (2017) 007} [\href{https://arxiv.org/abs/1703.08704}{{\ttfamily
  1703.08704}}].

\bibitem{Pate:2019lpp}
M.~Pate, A.-M.~Raclariu, A.~Strominger and E.Y.~Yuan, \emph{{Celestial operator
  products of gluons and gravitons}},
  \href{https://doi.org/10.1142/S0129055X21400031}{\emph{Rev. Math. Phys.}
  {\bfseries 33} (2021) 2140003}
  [\href{https://arxiv.org/abs/1910.07424}{{\ttfamily 1910.07424}}].

\bibitem{andrews1998theory}
G.E.~Andrews, \emph{The theory of partitions}, no.~2, Cambridge university
  press (1998).

\bibitem{chen2025}
B.~Chen, S.~He, P.~Mao and X.-C.~Mao, \emph{Gravitational path integral and
  partition function}, Work in progress, (2025).

\end{thebibliography}\endgroup

\end{document}